\pdfoutput=1
\documentclass[draft]{agujournal}
\draftfalse

\usepackage{epstopdf}

\journalname{JGR-Earth Surface}

\begin{document}
\title{The Cessation Threshold of Nonsuspended Sediment Transport Across Aeolian and Fluvial Environments}
\authors{Thomas P\"ahtz\affil{1,2} and Orencio Dur\'an\affil{3}}

\affiliation{1}{Institute of Port, Coastal and Offshore Engineering, Ocean College, Zhejiang University, 866 Yu Hang Tang Road, 310058 Hangzhou, China}
\affiliation{2}{State Key Laboratory of Satellite Ocean Environment Dynamics, Second Institute of Oceanography, 36 North Baochu Road, 310012 Hangzhou, China}
\affiliation{3}{Department of Ocean Engineering, Texas A\&M University, College Station, Texas 77843-3136, USA}

\correspondingauthor{Thomas P\"ahtz}{0012136@zju.edu.cn}

\begin{keypoints}
\item We propose an analytical model of the bulk sediment transport cessation threshold that reproduces measurements in air and viscous and turbulent liquids
\item Threshold of vanishing bulk transport is not a fluid or impact entrainment threshold but associated with sustaining particle-bed rebounds
\item We classify intermittent transport regimes using the energy criterion for incipient motion and our new insights on transport cessation
\end{keypoints}

\begin{abstract}
Using particle-scale simulations of non-suspended sediment transport for a large range of Newtonian fluids driving transport, including air and water, we determine the bulk transport cessation threshold $\Theta^r_t$ by extrapolating the transport load as a function of the dimensionless fluid shear stress (\textit{Shields number}) $\Theta$ to the vanishing transport limit. In this limit, the simulated steady states of continuous transport can be described by simple analytical model equations relating the average transport layer properties to the law of the wall flow velocity profile. We use this model to calculate $\Theta^r_t$ for arbitrary environments and derive a general Shields-like threshold diagram in which a Stokes-like number replaces the particle Reynolds number. Despite the simplicity of our hydrodynamic description, the predicted cessation threshold, both from the simulations and analytical model, quantitatively agrees with measurements for transport in air and viscous and turbulent liquids despite not being fitted to these measurements. We interpret the analytical model as a description of a continuous rebound motion of transported particles and thus $\Theta^r_t$ as the minimal fluid shear stress needed to compensate the average energy loss of transported particles during an average rebound at the bed surface. This interpretation, supported by simulations near $\Theta^r_t$, implies that entrainment mechanisms are needed to sustain transport above $\Theta^r_t$. While entrainment by turbulent events sustains intermittent transport, entrainment by particle-bed impacts sustains continuous transport. Combining our interpretations with the critical energy criterion for incipient motion by Valyrakis and coworkers, we put forward a new conceptual picture of sediment transport intermittency.
\end{abstract}

\section{Introduction} \label{Introduction}

One of the key interests in geomorphology is to predict the evolution of fluid-sheared surfaces, such as river beds, ocean floors, and wind-blown planetary surfaces \citep{Yalin77,Graf84,vanRijn93,Julien98,Garcia07,Bagnold41,Shao08,PyeTsoar09,Zheng09,Bourkeetal10,Duranetal11,Merrison12,Koketal12,Rasmussenetal15,Shaoetal15,Valanceetal15}. The ability to do so requires answering two basic questions: what is the shear stress $\tau$ a stream of Newtonian fluid needs to exert onto a bed composed of loose sediment to initiate transport and what is the shear stress below which transport once initiated ceases. As the fluid force acting on sediment particles ($\sim\tau d^2$) must overcome their buoyant weight [$\sim(\rho_p-\rho_f)gd^3$], one often recasts these problems in terms of the dimensionless fluid shear stress (\textit{Shields number}) $\Theta=\tau/[(\rho_p-\rho_f)gd]$, where $\rho_p$ ($\rho_f$) is the particle (fluid) density, $g$ the gravitational constant, and $d$ the characteristic particle diameter.

The definitions of both the Shields number at the initiation threshold (also known as \textit{static threshold}, \textit{fluid threshold}, \textit{entrainment threshold}, \textit{erosion threshold}, and \textit{incipient motion threshold}) and cessation threshold (also known as `dynamic threshold' and `impact threshold'), however, are not trivial as there is no obvious criterion distinguishing transport from no transport. The problem is that motion of individual sediment particles, however slow and short-lasting it may be, can occur at any nonzero $\Theta$ due to subsurface creep \citep{Houssaisetal15}, turbulent events, and the spatial and temporal scale of observation \citep{Paintal71,LavelleMofjeld87,Laursenetal99,Salimetal17}. For these and other reasons, various criteria for \textit{significant} transport have been proposed in the literature to visually identify the threshold Shields numbers. In general, one can distinguish criteria based on critical amounts of individually moving particles, such as the \textit{weak} and \textit{medium} motion criteria by \citet{Kramer35}, from criteria based on the occurrence of, or a critical amount of, bulk transport, defined as a comparably long-lasting collective particle motion, such as the \textit{general} motion criterion by \citet{Kramer35}. The former criteria are typically used for turbulent fluvial flows \citep[][and references therein]{Milleretal77,BuffingtonMontgomery97,Paphitis01}, whereas the latter criteria are preferred for laminar fluvial \citep{YalinKarahan79,Govers87} and turbulent aeolian flows \citep{Bagnold36,Bagnold37,Bagnold41,Chepil45,LylesKrauss71,Iversenetal76,Greeleyetal76,IversenWhite82,IversenRasmussen94,Dongetal03b,Burretal15,Carneiroetal15,Webbetal16} because bulk transport can be easily visually identified in these environments: by the formation of \textit{grain carpets} in laminar fluvial flows and by very large particle hops in turbulent aeolian flows. A further situation sometimes studied in the laboratory is the beginning motion of a single particle on top of a prearranged substrate \citep{FentonAbbott77,Charruetal07,Diplasetal08,Celiketal10,Valyrakisetal10,Valyrakisetal11,Valyrakisetal13,AgudoWierschem12,Agudoetal14,Agudoetal17,Agudoetal18,Kudrollietal16,DeskosDiplas18}. Moreover, it is worth highlighting that \citet{Salevanetal17} proposed a fundamentally different criterion for the onset of significant motion based on analyzing the particle velocity distribution of all particles, including those nearly static ones that belong to the bed surface.

The main problem with all these various criteria is the ambiguity of the associated thresholds. What exactly are their different meanings in natural settings, and how relevant are they for the evolution of fluid-sheared surfaces? An alternative definition of the threshold, which we focus on in this study, treats this problem backward. Because predicting the evolution of fluid-sheared surfaces usually requires knowledge of the amount of nonsuspended sediment carried along the surface, one defines a threshold Shields number $\Theta^t_r$ indirectly through an equation relating the transport rate $Q$ of nonsuspended sediment to $\Theta$, such as
\begin{linenomath*}
\begin{equation}
 Q\propto\rho_pd(\Theta-\Theta^t_r)\overline{v_x}, \label{ThresholdDef}
\end{equation}
\end{linenomath*}
where $\overline{v_x}$ is the average particle velocity in the flow direction. Equation~(\ref{ThresholdDef}) has been used to obtain $\Theta^t_r$ via extrapolation of paired measurements of $\Theta$ and $Q$, or an indicator of $Q$, to vanishing transport or a small reference value (the \textit{reference method}) \citep[][and references therein]{Shields36,Milleretal77,ParkerKlingeman82,BuffingtonMontgomery97,Paphitis01,Charruetal04,Cliftonetal06,Ouriemietal07,Lobkovskyetal08,Hongetal15,Creysselsetal09,BarchynHugenholtz11,Hoetal11,Ho12,Lietal14}. Because equation~(\ref{ThresholdDef}) predicts $Q(\Theta^r_t)=0$ even though transport never truly vanishes for $\Theta>0$, it is apparent that $Q$ in equation~(\ref{ThresholdDef}), if taken at face value, cannot include all occurring nonsuspended transport provided that $\Theta^r_t$ has a physical meaning beyond that of a fit parameter (the working hypothesis of this study). Given that equation~(\ref{ThresholdDef}) stays representative for the large class of Bagnoldian transport relations \citep[e.g.,][]{Bagnold56,Bagnold73,AbrahamsGao06,Charru06,AliDey17,Owen64,Sorensen91,Sorensen04,MartinKok17}, which are usually derived assuming a mean turbulent flow without fluctuations, one may conclude that $Q$ in equation~(\ref{ThresholdDef}) does not include individual sediment particle motion caused by sporadic turbulent events and thus that the threshold $\Theta^r_t$ is associated with vanishing bulk transport, at least in the Bagnoldian sense. In fact, Bagnoldian transport relations (not to be confused with Bagnold's stream power relations \citep{Bagnold66,Bagnold77,Bagnold80}) are physically based on Bagnold's assumption, recently numerically validated \citep{Duranetal12}, that the particle-flow feedback induced by continuous bulk transport reduces the dimensionless fluid shear stress at the bed surface to a value close to $\Theta^r_t$. This class of relations has been shown to describe continuous nonsuspended bulk transport in direct transport simulations for a large range of the particle-fluid-density ratio $s=\rho_p/\rho_f$ and Galileo number $\mathrm{Ga}=\sqrt{(s-1)gd^3}/\nu$ \citep{Duranetal11,Duranetal12,Duranetal14a,Charruetal16}, where $\nu$ is the kinematic fluid viscosity. In this study, continuous transport is defined as transport that never stops for any system size that is larger than a certain cutoff size below which finite size effects begin to play a role.

In the fluvial transport community \citep[][and references therein]{Milleretal77,BuffingtonMontgomery97,Paphitis01} (and in parts of the aeolian transport community \citep{Shao08,Raffaeleetal16,Raffaeleetal18}), $\Theta^r_t$ is currently treated as an initiation threshold. One may therefore wonder whether the paired data $Q(\Theta)$ used to obtain $\Theta^r_t$ through the reference method depend on the experimental protocol. In fact, in fluvial environments, these data are often measured using an initiation protocol (i.e., successively incrementing $\Theta$ starting from a static bed), which may yield results different from a cessation protocol (i.e., successively decrementing $\Theta$ starting from a mobile bed) because of changes in the structure and size distribution of bed surfaces in response to transport events \citep[e.g.,][]{Turowskietal11}. Apart from these effects, the direct transport simulations of fluvial sediment transport by \citet{Maurinetal15} suggest that initiation and cessation protocols yield similar results for realistic settings (i.e., when turbulent fluctuations around the mean turbulent flow are considered in the simulations) as the data for both protocols collapse on a master curve $Q(\Theta)$ of a structure consistent with equation~(\ref{ThresholdDef}) (Figure~S1 in the supporting information). In contrast, for the unrealistic absence of turbulent fluctuations, the master curve $Q(\Theta)$ is only obtained when the cessation protocol is used (Figure~S1). These two numerical findings also suggest that, if and only if fluvial transport has reached a state characterized by the master curve $Q(\Theta)$, it becomes insensitive to the history of how this state has been reached (e.g., by a turbulent fluctuation event or an initially mobile bed). Based on Figure~S1, we assume that initiation and cessation protocols are similar, but we acknowledge that the Figure~S1 simulations do not consider changes in bed surface structure and grain size that may occur during transport events; accounting for those effects might cause the initiation and cessation protocols to yield somewhat different results.

Put together, the findings that the functional structure of equation~(\ref{ThresholdDef}) may be insensitive to fluid properties, turbulent fluctuations, and the data acquisition protocol question the current paradigm that different physical mechanisms control $\Theta^r_t$ in different environments, such as fluid entrainment in fluvial environments \citep{Paphitis01} and entrainment by particle-bed impacts in aeolian environments (`splash') \citep{Koketal12}. Indeed, here we show that there is a universal physical meaning of $\Theta^r_t$ and develop a relatively simple general model predicting $\Theta^r_t$ across environments in agreement with available measurements.

The paper is organized as follows. In section~\ref{Simulations}, we introduce the numerical model of sediment transport in a Newtonian fluid by \citet{Duranetal12}, which we use to simulate nonsuspended sediment transport for a large range of the density ratio ($s\in[1.1,2035]$) and Galileo number ($\mathrm{Ga}\in[0.1,156]$) and subsequently determine $\Theta^r_t$ through a cessation protocol. The simulated range includes various \textit{saltation} transport conditions (typical for, but not limited to, aeolian environments), in which particles predominantly move without being in contact with other particles (e.g., hopping, Movie~S1), and various \textit{bedload} transport conditions (typical for, but not limited to, fluvial environments), in which intergranular contacts (e.g., sliding and rolling) significantly affect the particle motion (Movie~S2). (Note that Movie~S2 looks different from natural fluvial bedload transport because turbulent fluctuations around the mean turbulent flow are not considered in the simulations.) In section~\ref{Results}, we validate the numerical simulations with experimental transport rate and threshold data corresponding to aeolian transport on Earth and viscous and turbulent fluvial transport. We then show that the average properties of the simulated transport states can be described by a set of equations that can be used to analytically determine $\Theta^r_t$ for arbitrary environments. In section~\ref{Discussion}, we conceptually compare this analytical model with other threshold models from the literature and discuss our results in the context of bed sediment entrainment, incipient sediment motion, and transport intermittency. Finally, we draw conclusions in section~\ref{Conclusions}.

\section{Numerical Model Description} \label{Simulations}
The numerical model of sediment transport in a Newtonian fluid by \citet{Duranetal12} belongs to a new generation of sophisticated grain-scale models of sediment transport \citep{Carneiroetal11,Duranetal11,Duranetal12,Carneiroetal13,Jietal13,Duranetal14a,Duranetal14b,KidanemariamUhlmann14a,KidanemariamUhlmann14b,KidanemariamUhlmann17,Schmeeckle14,Vowinckeletal14,Vowinckeletal16,ArollaDesjardins15,Pahtzetal15a,Pahtzetal15b,Carneiroetal15,Clarketal15,Clarketal17,Derksen15,Maurinetal15,Maurinetal16,Maurinetal18,FinnLi16,Finnetal16,SunXiao16,ElghannayTafti17a,ElghannayTafti17b,Gonzalezetal17,PahtzDuran17,Seiletal18}. It couples a discrete element method for the particle motion with a continuum Reynolds-averaged description of hydrodynamics, which means that it neglects turbulent fluctuations around the mean turbulent flow. It simulates the translational and rotational dynamics of $\approx15,000$ spheres, including $>10$ layers of bed particles (more than sufficient to completely dissipate the energy of particles impacting the bed surface), with diameters $d_p$ evenly distributed within two sizes ($0.8d$ and $1.2d$) in a quasi-2-D, vertically infinite domain of length $1181d$. Periodic boundary conditions are imposed along the flow direction, while the bottommost layer of particles is glued to a bottom wall. The particle contact model considers normal repulsion, energy dissipation, and tangential friction, where the magnitude of the tangential friction force relative to the normal contact force is limited through a Coulomb friction criterion. The Reynolds-averaged Navier-Stokes equations are applied to an inner turbulent boundary layer of infinite size, which means that the flow depth of fluvial flows is assumed to be much larger than the thickness of the bedload transport layer. These equations are combined with an improved mixing length approximation that ensures a smooth hydrodynamic transition from high to low particle concentration at the bed surface and quantitatively reproduces the law of the wall flow velocity profile in the absence of transport. The model considers the gravity, buoyancy, added-mass, and fluid drag force acting on particles. However, cohesive and higher-order fluid forces, such as the lift force and hindrance effect on the drag force are neglected. We refer the reader to the original publication \citep{Duranetal12} for further details (note that we recently corrected slight inaccuracies in the original model \citep{PahtzDuran17}).

\subsection{Sensitivity to Contact Model Parameters (Lubrication Forces)} \label{Lubrication}
All results presented in this study for the bedload transport regime do not significantly depend on contact parameters, such as the normal restitution coefficient $e$ and tangential contact friction coefficient $\mu_c$. For example, the transport rate $Q$ and threshold $\Theta^r_t$ are nearly the same for bedload transport simulations with $e=0.9$ and $e=0.01$, which is consistent with previous studies \citep{DrakeCalantoni01,Maurinetal15,ElghannayTafti17b,PahtzDuran17}. The probable reason is that nearly maximal dissipative normal collisions ($e=0.01$) do not prevent particles from rebounding at the bed surface (Movie~S2), even for aeolian saltation transport (Movie~S3), because the contact plane is usually inclined against the horizontal plane for particle-bed collisions as the bed is not flat. However, note that the contact parameters $e$ and $\mu_c$ significantly affect the granular flow rheology below the bed surface, most notably the local energy dissipation rate. In particular, the number of bed layers needed to dissipate the energy that particles lose when rebounding at the bed surface increases with $e$. The insensitivity of the transport layer to contact parameters implies that any dissipative short-range force that is proportional to the relative velocity of two interacting particles, such as the lubrication force \citep{SimeonovCalantoni12}, does not significantly influence average bedload transport characteristics as the effect of such forces can be incorporated in $e$ and $\mu_c$ \citep{Gondretetal02,YangHunt06,Schmeeckle14,Maurinetal15}. It also challenges recent analytical models of sediment transport that assumed a flat bed and thus a crucial role of lubrication forces \citep{Berzietal16,Berzietal17}. Note that, for all simulation data shown in this manuscript, $e=0.9$ (except in Movies~S2 and S3) and $\mu_c=0.5$, which are values typical for dry quartz sand.

\subsection{Computation of Average Quantities From the Simulation Data}

\subsubsection{Local Average Over Particles}
We define a Cartesian coordinate system $\mathbf{x}=(x,y,z)$, where $x$ is the horizontal coordinate in the flow direction parallel to the sediment bed, $z$ the vertical coordinate normal to the bed, and $y$ the lateral coordinate. Using this coordinate system, we compute the local mass-weighted ensemble average of a quantity $A$ through \citep{Pahtzetal15a}
\begin{linenomath*}
\begin{equation}
 \langle A\rangle=\frac{1}{\rho}\overline{\sum_nm^nA^n\delta(\mathbf{x}-\mathbf{x}^n)}^E,
\end{equation}
\end{linenomath*}
where $\mathbf{x}^{n}$ ($m^n$) is the location (mass) of particle $n$, $\rho=\overline{\sum_nm^n\delta(\mathbf{x}-\mathbf{x}^n)}^E$ the particle mass density, $\delta$ the delta distribution, and the overbar with superscript \textit{E} denotes the ensemble average, which we calculate for our simulated steady systems through time averaging over the simulation duration (ergodic hypothesis). The $\delta$ kernels are further coarse grained through spatial averaging over a discretization volume of size $1181d\times1d\times\Delta z$, where $\Delta z$ varies between $0.05d$ in dense and dilute regions and larger values in rarefied regions. Henceforth, the $\delta$ symbol should thus be interpreted as the associated coarse-graining function.

\subsubsection{Particle Stress Tensor}
The particle stress tensor $P_{ij}$ is composed of a transport contribution (superscript \textit{t}) and a contact contribution (superscript \textit{c}) and computed through \citep{Pahtzetal15a}
\begin{linenomath*}
\begin{subequations}
\begin{eqnarray}
 P_{ij}&=&P^t_{ij}+P^c_{ij} \label{ParticleStress}, \\
 P^t_{ij}&=&\rho(\langle v_iv_j\rangle-\langle v_i\rangle\langle v_j\rangle), \label{Pt} \\
 P^c_{ij}&=&\frac{1}{2}\overline{\sum_{mn}F^{mn}_j(x^m_i-x^n_i)\int\limits_0^1\delta(\mathbf{x}-((\mathbf{x}^m-\mathbf{x}^n)s^\prime+\mathbf{x}^n))\mathrm{d}s^\prime}^E,
\end{eqnarray}
\end{subequations}
\end{linenomath*}
where $s^\prime$ is a dummy variable and $\mathbf{F}^{mn}$ the contact force applied by particle $n$ on particle $m$ ($\mathbf{F}^{mm}=0$). We confirmed that this manner of computation is equivalent to computing $P_{zx}$ and $P_{zz}$ indirectly through the momentum balance. For steady, homogeneous transport conditions ($\partial/\partial_t=\partial/\partial_x=\partial/\partial_y=0$), the mass and momentum balance read \citep{Pahtzetal15a}
\begin{linenomath*}
\begin{eqnarray}
 \text{Mass balance:}&&\quad\;\;\langle v_z\rangle=0, \label{MassBalance} \\
 \text{Momentum balance:}&&\mathrm{d}P_{zi}/\mathrm{d}z=\rho\langle a_i\rangle, \label{MomentumBalance}
\end{eqnarray}
\end{linenomath*}
where $\mathbf{a}$ is the particle acceleration due to noncontact forces.

\subsubsection{Bed Surface Elevation}
We compute the bed surface $z_r$ (i.e., the lower limit of the transport layer) as the elevation where the production rate $P_{zz}\mathrm{d}\langle v_x\rangle/\mathrm{d}z$ of the cross-correlation fluctuation energy density $-\rho\langle(v_x-\langle v_x\rangle)v_z\rangle$ is maximal \citep{PahtzDuran17}. Indeed, $z_r$ is a measure for the effective location of energetic particle-bed rebounds, which are the main reason for the production of $-\rho\langle(v_x-\langle v_x\rangle)v_z\rangle$ because they effectively convert the horizontal momentum of descending particles into the vertical momentum of ascending particles. Consistently, $\mathrm{d}\langle v_x\rangle/\mathrm{d}z$ always peaks near the bed surface due to an exponential decay of $\langle v_x\rangle$ within the sediment bed \citep{PahtzDuran17}.

\subsubsection{Transport Layer Average} \label{TransportLayerAverage}
The average over the transport layer of a quantity $A$ is represented by an overbar and defined as
\begin{linenomath*}
\begin{equation}
 \overline{A}=\frac{1}{M_r}\int\limits_{z_r}^\infty\rho\langle A\rangle\mathrm{d}z, \label{HeightAverage}
\end{equation}
\end{linenomath*}
where $M_r=\int_{z_r}^\infty\rho\mathrm{d}z$ is the transport load (i.e., the mass of transported particles per unit area). Note that $M_r=Q_r/\overline{v_x}$, where $Q_r=\int_{z_r}^\infty\rho\langle v_x\rangle\mathrm{d}z$ is the rate of sediment transport above the bed surface, which is slightly smaller than the total transport rate $Q=\int_{-\infty}^\infty\rho\langle v_x\rangle\mathrm{d}z$.

\section{Results} \label{Results}
This section is separated into three subsections: In section~\ref{ModelValidation}, we validate the numerical model with experimental data, including measurements of the transport rate $Q$ and threshold $\Theta^r_t$. In section~\ref{AnalyticalModel}, we derive the equations that describe the simulated steady, continuous transport states, use them to predict $\Theta^r_t$ in arbitrary environments, and validate these analytical predictions with available experimental data. In section~\ref{ThresholdInterpretation}, we explain why we interpret $\Theta^r_t$ as the minimal fluid shear stress needed to compensate the average energy loss of rebounding particles by the fluid drag acceleration during particle trajectories. Finally, in section~\ref{ModShields}, we analyze the predictions of the analytical model and present a general threshold diagram.

\subsection{Model Validation} \label{ModelValidation}
\subsubsection{Validation With Nonthreshold Measurements}
\begin{figure}[htb]
 \begin{center}
  \includegraphics[width=1.0\columnwidth]{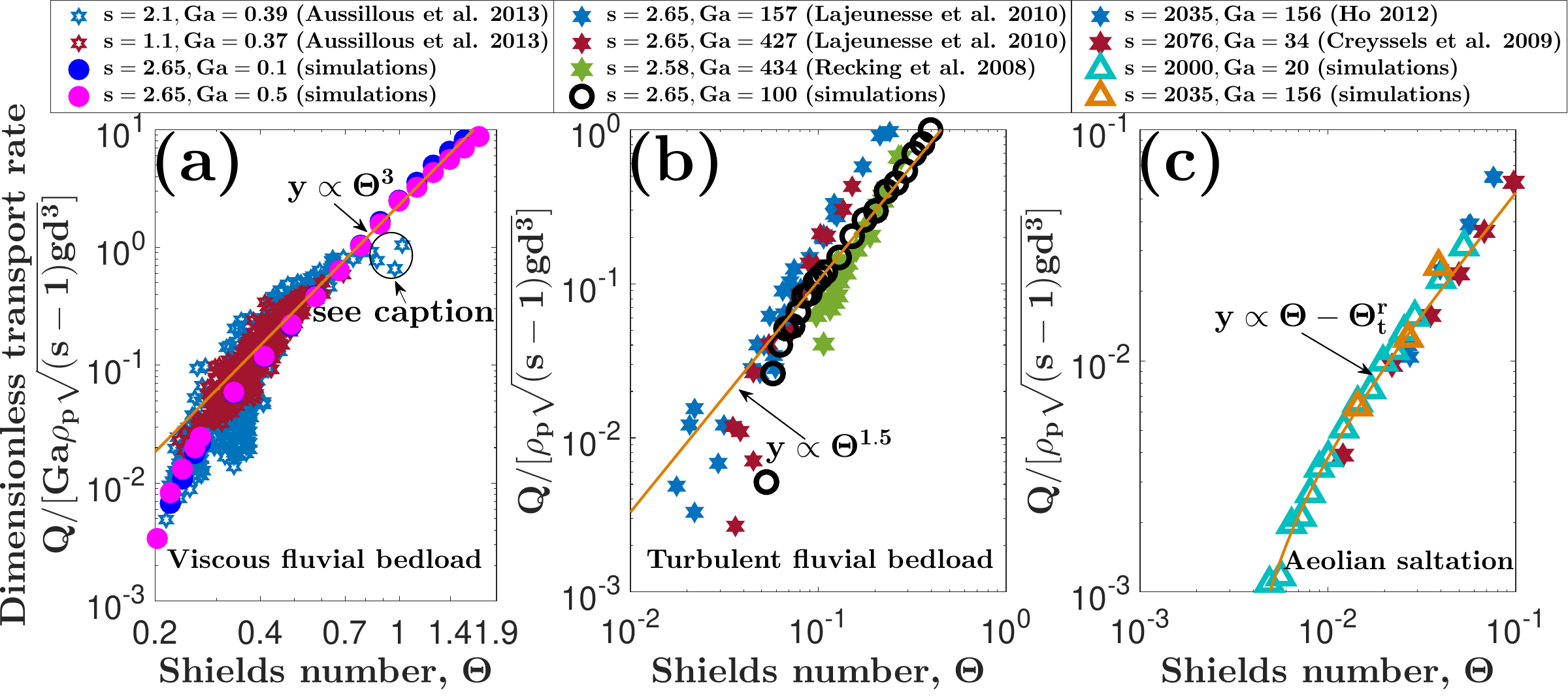}
 \end{center}
 \caption{{\bf Numerical model versus measurements.} Comparison between simulations and measurements \citep{Reckingetal08b,Lajeunesseetal10,Aussillousetal13,Creysselsetal09,Ho12} of the dimensionless sediment transport rate as a function of the Shields number $\Theta$ for (a) viscous fluvial bedload, (b) turbulent fluvial bedload, and (c) aeolian saltation transport. The nondimensionalization of the sediment transport rate $Q$ in (a) is different from the one in (b) and (c) in order to account for the proportionality of $Q$ with $\mathrm{Ga}$ for viscous bedload transport \citep{KidanemariamUhlmann14b}. The saturation of $Q$ when $\Theta$ is large, present in some viscous bedload measurements (a), is caused by more and more particles moving near the top of the experimental facility with increasing $\Theta$ \citep{Aussillousetal13}, where the flow velocity does not follow the laminar velocity profile due to the nonslip condition at the upper boundary. In contrast, there is no upper boundary in the direct transport simulations.}
\label{ModelComparisons1}
\end{figure}

\begin{figure}[htb]
 \begin{center}
  \includegraphics[width=1.0\columnwidth]{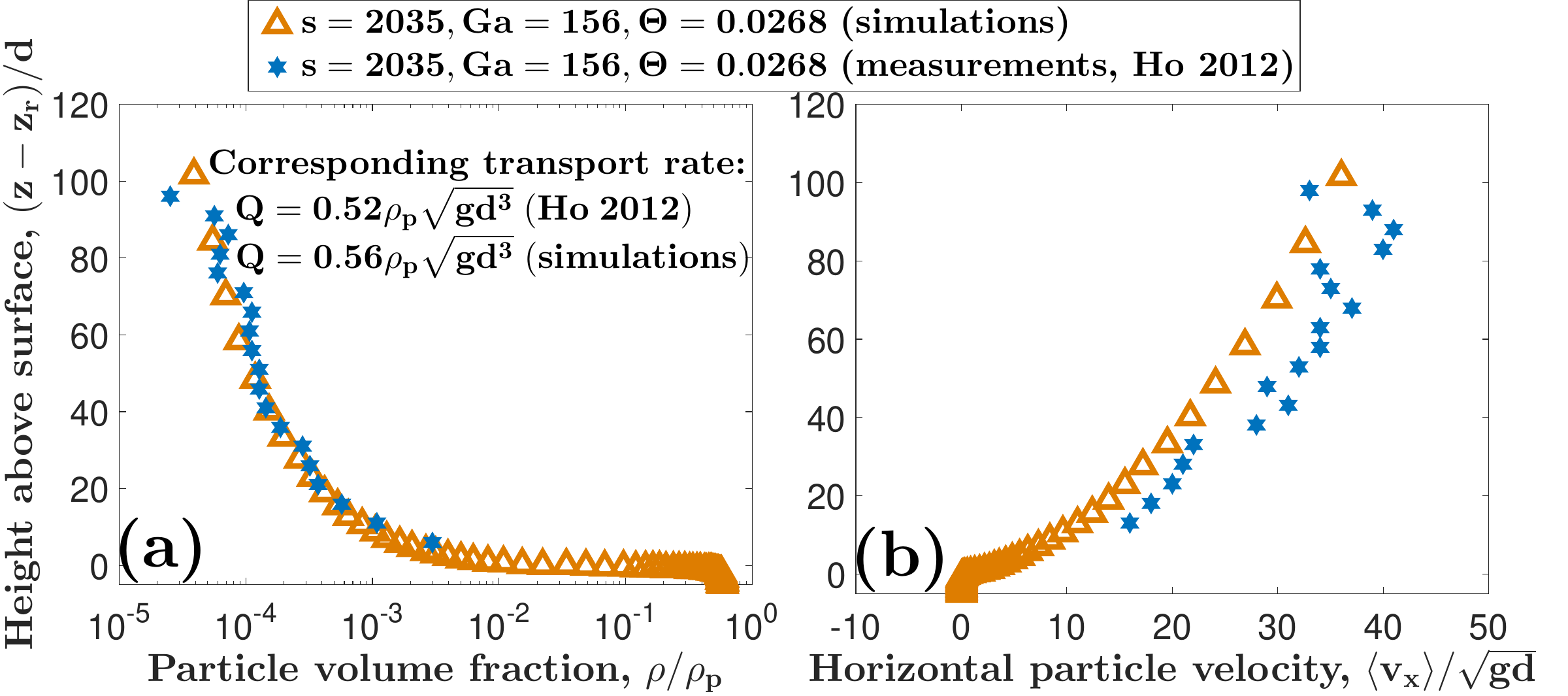}
 \end{center}
 \caption{{\bf Numerical model versus measurements.} Comparison between simulations and aeolian saltation transport measurements \citep{Ho12} of the vertical profiles relative to the bed surface elevation $z_r$ of (a) the particle volume fraction $\rho/\rho_p$ and (b) the average particle velocity $\langle v_x\rangle$ in the flow direction.}
\label{ModelComparisons2}
\end{figure}

The model by \citet{Duranetal12} has previously been shown to reproduce many observations concerning bedform formation \citep{Duranetal14b} and the quality of viscous and turbulent sediment transport in air and liquids \citep{Duranetal11,Duranetal12,Duranetal14a}. Here we show further quantitative model tests in Figures~\ref{ModelComparisons1} and \ref{ModelComparisons2}.
Figure~\ref{ModelComparisons1} compares model simulations against measurements of the dimensionless sediment transport rate for three very different environmental conditions that are characterized by varying combinations of $s$ and $\mathrm{Ga}$: viscous fluvial bedload (Figure~\ref{ModelComparisons1}a), turbulent fluvial bedload (Figure~\ref{ModelComparisons1}b), and aeolian saltation transport (Figure~\ref{ModelComparisons1}c). This comparison reveals that the simulations do not only reproduce the very different (asymptotic) scaling laws but are also quantitatively consistent with measurements of $Q$. Figure~\ref{ModelComparisons2} shows that the simulations are also quantitatively consistent with aeolian saltation transport measurements of the vertical profiles of the particle volume fraction $\rho/\rho_p$ and average horizontal particle velocity $\langle v_x\rangle$. A quantitative comparison against measurements of the vertical profiles of the average fluid and particle velocity in viscous fluvial bedload transport was already shown in Figure~6 of \citet{Duranetal14a}.

\subsubsection{Validation With Threshold Measurements}
The method that we use to obtain $\Theta^r_t$ from the steady, continuous transport simulation data is adopted from \citet{Duranetal11,Duranetal12}. We define a flux-weighted height average of the horizontal particle velocity: $\overline{v_x}^q=\int_{-\infty}^\infty\rho\langle v_x^2\rangle\mathrm{d}z/Q$ and linearly extrapolate $M=Q/\overline{v_x}^q$ as a function of the Shields number $\Theta$ to $M=0$, consistent with equation~(\ref{ThresholdDef}). In section~\ref{SteadyState}, we validate this extrapolation procedure using simulations of the temporal decay of $M$ across the threshold (an interpolation procedure that accounts for transport intermittency near $\Theta^r_t$). In particular, it turns out that $M>0$ even when $\Theta\leq\Theta^r_t$, consistent with subsurface creeping \citep{Houssaisetal15}, as mentioned in section~\ref{Introduction}. However, the associated values of $M$ are so small that they do not matter for the extrapolation procedure. The quantity $M$ is approximately proportional to $M_r=Q_r/\overline{v_x}$ (section~\ref{TransportLayerAverage}) and thus a measure for the transport load because $Q_r\simeq Q$ and $\overline{v_x}\propto\overline{v_x}^q$ are approximately obeyed. However, the data $M(\Theta)$ are significantly smoother than $M_r(\Theta)$, which allows a more accurate determination of $\Theta^r_t$. Note that, for most conditions, it does not matter much for the value of $\Theta^r_t$ how many points are considered for the linear extrapolation. However, for viscous bedload transport, there is a change in the slope of $M(\Theta)$ not far from $\Theta^r_t$ \citep{Duranetal14a}. For such cases, we therefore use only points sufficiently near $\Theta^r_t$ for the extrapolation.

We compare the so obtained values of $\Theta^r_t$ with experimental data from three different conditions:

 \noindent \textit{Bedload transport in viscous liquids:} We use the threshold measurements by \citet{YalinKarahan79}, who identified the onset of bulk bedload transport (\textit{grain carpets}). According to our direct transport simulations, the associated threshold Shields number is nearly equivalent to $\Theta^r_t$ for viscous bedload transport conditions. We decided against data sets from other authors that measured the threshold via extrapolation \citep[e.g.,][]{Charruetal04,Ouriemietal07,Lobkovskyetal08,Hongetal15} due to the above mentioned slope change of $M(\Theta)$ near $\Theta^r_t$ for viscous bedload transport.

\noindent \textit{Bedload transport in turbulent liquids:} We carry out a compilation of measurements of $\Theta^r_t$, obtained via extrapolating sediment transport rate relations like equation~(\ref{ThresholdDef}) to vanishing transport or a small reference value (i.e., the reference method), provided and referenced in the review by \citet{BuffingtonMontgomery97}. We assume that the various reference value definitions (including vanishing transport) yield similar results and that the transport rate data $Q(\Theta)$ used for the extrapolation do not significantly depend on the experimental protocol (measuring $Q$ via successively incrementing versus decrementing $\Theta$) for reasons explained in section~\ref{Introduction}.

\noindent \textit{Saltation transport in air:} We use the paired measurements of $Q$ and $\Theta$ reported by \citet{Creysselsetal09} and \citet{Hoetal11} to obtain $\Theta^r_t$ via linear extrapolation to vanishing transport. However, we only use measurements with $\Theta<0.06$ because this value marks the beginning of the nonlinear \textit{Bagnold regime} in direct transport simulations \citep[][their Figure~27]{Duranetal11} (a paper on this issue is in preparation). We also choose visual measurements of the \textit{impact threshold} by \citet{Bagnold37}, \citet{Chepil45}, and \citet{Carneiroetal15} for the comparison (for the data in Figure~6 of \citet{Carneiroetal15}, we estimated a slightly smaller value $0.0049$ than the value $0.0053$ these authors reported). As we discuss in section~\ref{Intermittency}, these measurements may be close to $\Theta^r_t$.

\begin{figure}[htb]
 \begin{center}
	\includegraphics[width=1.0\columnwidth]{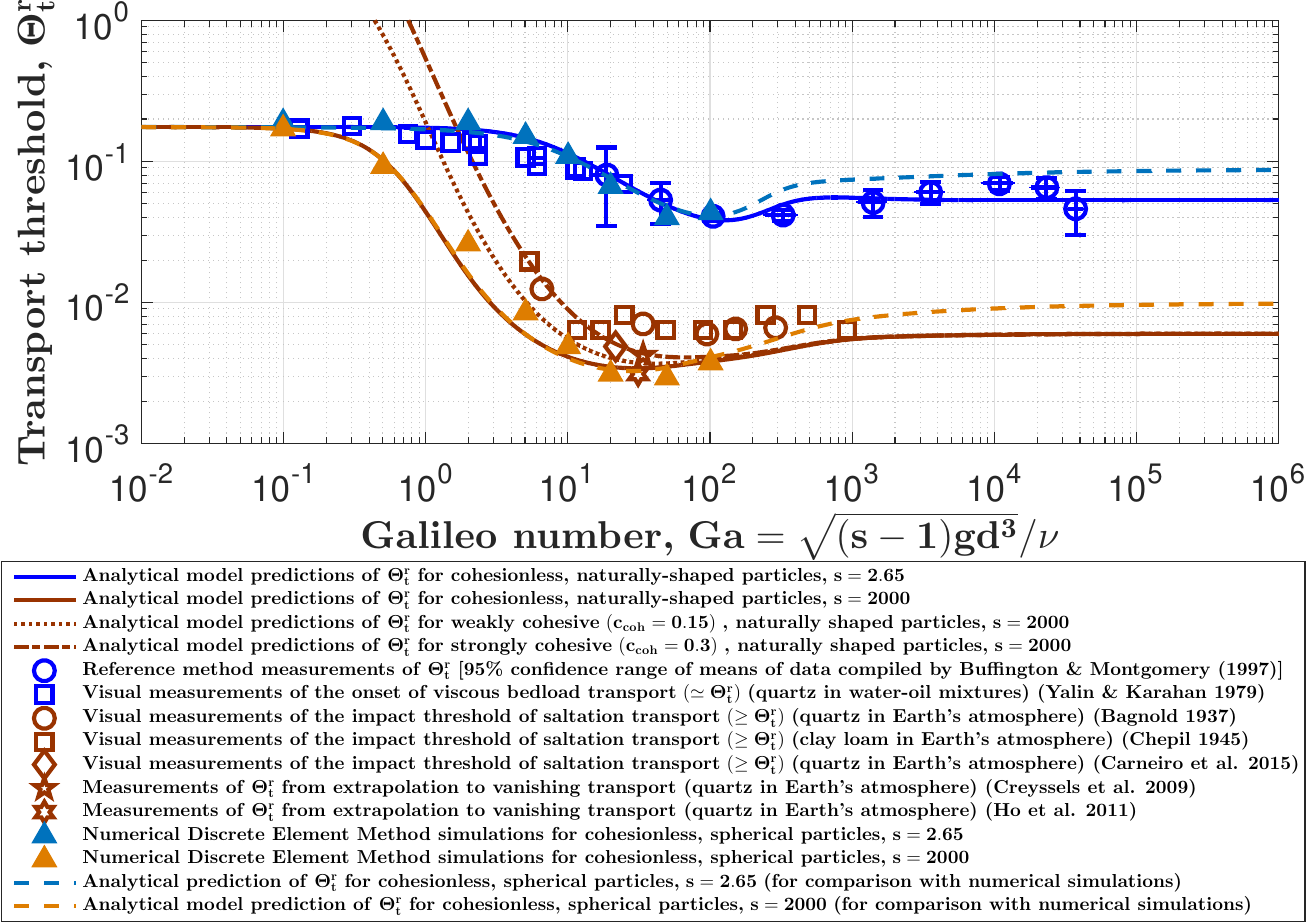}
 \end{center}
 \caption{{\bf Validation of numerical and analytical threshold predictions.} Threshold Shields number $\Theta^r_t$ versus Galileo number $\mathrm{Ga}$ for simulated (filled symbols) and measured (open symbols) values in water and water-oil mixtures \citep{YalinKarahan79,BuffingtonMontgomery97} (blue) and Earth's atmosphere \citep{Bagnold37,Chepil45,Creysselsetal09,Hoetal11,Carneiroetal15} (brown). Predictions from the analytical model are shown as lines.}
 \label{ModelvsMeasurements}
\end{figure}

Figure~\ref{ModelvsMeasurements} shows that the values of $\Theta^r_t$ obtained from our direct transport simulations (filled symbols) are consistent with the measurements (open symbols) except for the aeolian impact threshold measurements at small Galileo number $\mathrm{Ga}$.
These measurements were obtained using small particles ($d<100\;\mu$m), which are largely affected by cohesive interparticle forces \citep{Bagnold37,Chepil45} neglected in the simulations. The lines in Figure~\ref{ModelvsMeasurements} correspond to the predictions from the analytical threshold model, which is presented in the following.

\subsection{Analytical Threshold Model} \label{AnalyticalModel}
For pedagogical reasons, we first summarize the analytical model and the physical meaning of each model equation and present an in-detail justification afterward. The main assumption behind the analytical model is that, for any value of the Shields number $\Theta>\Theta^r_t$, there is a unique equilibrium (i.e., steady) transport state that obeys equation~(\ref{ThresholdDef}). Under this assumption, the threshold $\Theta^r_t$ can be obtained by extrapolating this transport state to vanishing transport. We support and discuss this assumption in the context of previous results \citep{Carneiroetal11,Clarketal15} in section~\ref{SteadyState}. 

\begin{figure}[htb]
 \begin{center}
  \includegraphics[width=1.0\columnwidth]{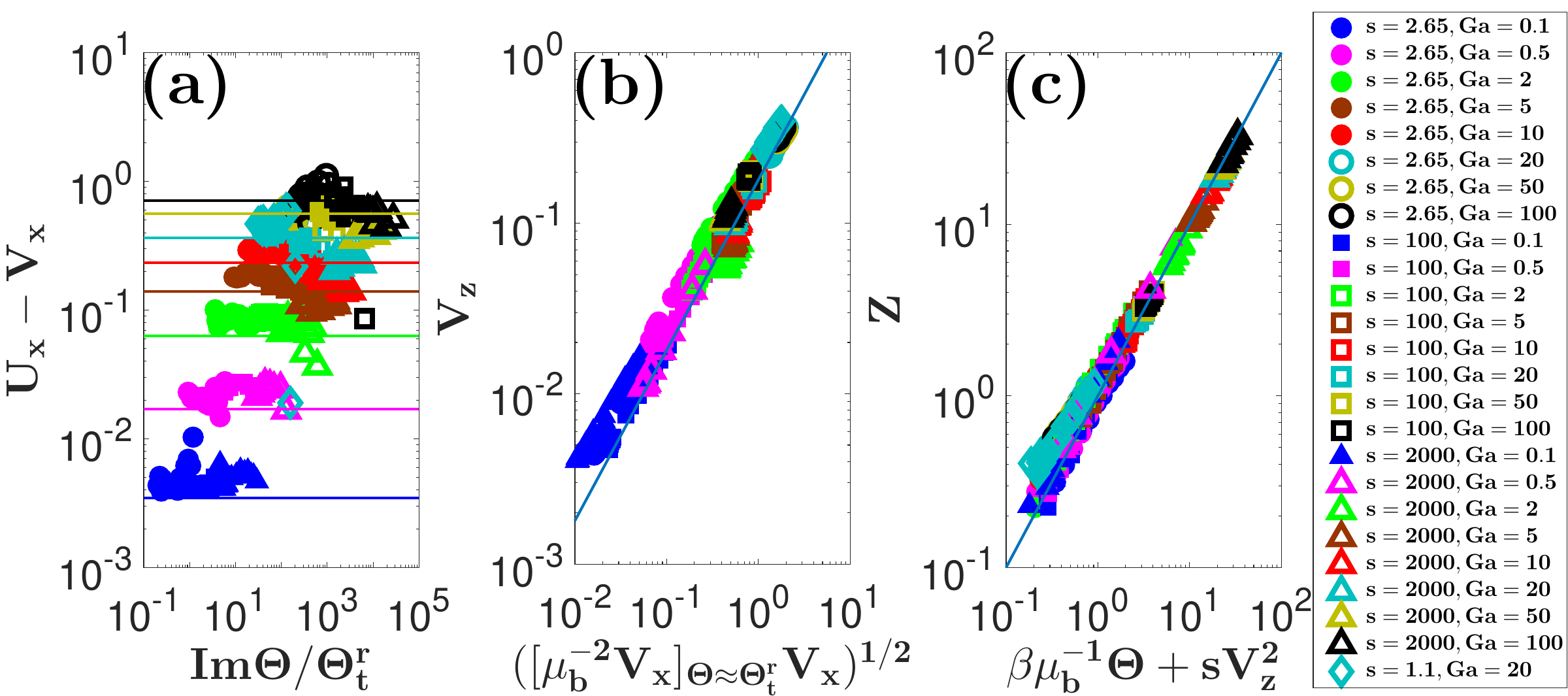}
 \end{center}
 \caption{{\bf Validation of equations~(\ref{mub}-\ref{Meanz}).} (a) Dimensionless fluid-particle-velocity difference $U_x-V_x$ versus product of \textit{impact number} $\mathrm{Im}=\mathrm{Ga}\sqrt{s+0.5}$ \citep{PahtzDuran17} and $\Theta/\Theta^r_t$. (b) Dimensionless vertical velocity $V_z$ versus $\sqrt{[\mu_b^{-2}V_x]_{\Theta\approx\Theta^r_t}V_x}$. (c) Dimensionless characteristic transport layer thickness $Z$ versus $\beta\mu_b^{-1}\Theta+sV_z^2$. Symbols correspond to data from our cohesionless transport simulations ($c_{\mathrm{coh}}=0$) for varying $s$, $\mathrm{Ga}$, and $\Theta$. The solid lines correspond to predictions from the analytical model. In particular, the colored solid lines in (a) correspond to the same value of $\mathrm{Ga}$ as the likewise colored symbols. Simulated values of the bed friction coefficient $\mu_b$, which slightly vary around the simulation mean $\mu_b=\mu^o_b$ (Figure~S2), are used in (b) and (c), and $\mu_b=\mu^o_b$ in (a).}
\label{AnalyticalModelValidation1}
\end{figure}

\begin{figure}[htb]
 \begin{center}
  \includegraphics[width=1.0\columnwidth]{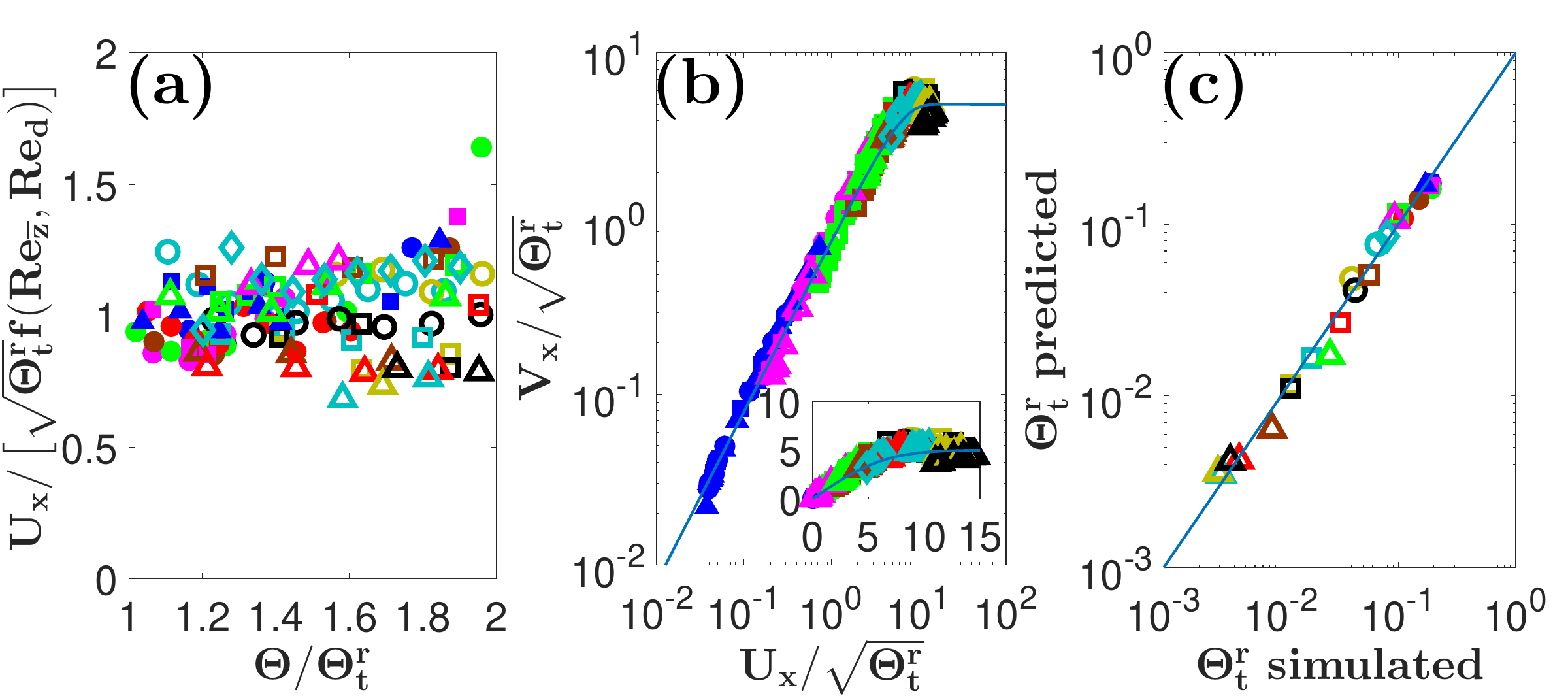}
 \end{center}
 \caption{{\bf Validation of equations~(\ref{MeanU}) and (\ref{VU}) and the prediction of $\Theta^r_t$ from the full model.} (a) $U_x/[\sqrt{\Theta^r_t}f(\mathrm{Re}_{\overline{z}},\mathrm{Re}_d)]$ versus $\Theta/\Theta^r_t$. (b) $V_x/\sqrt{\Theta^r_t}$ versus $U_x/\sqrt{\Theta^r_t}$. Symbols correspond to data from our cohesionless transport simulations ($c_{\mathrm{coh}}=0$) for varying $s$, $\mathrm{Ga}$, and $\Theta$ near the threshold $\Theta^r_t$. The solid lines correspond to predictions from the analytical model. (c) Predicted versus simulated values of $\Theta^r_t$ for varying $s$ and $\mathrm{Ga}$. For symbol legend, see Figure~\ref{AnalyticalModelValidation1}.}
\label{AnalyticalModelValidation2}
\end{figure}

We find that steady, continuous transport is described by three equations relating the average horizontal fluid velocity $U_x=\overline{u_x}/\sqrt{(s-1)gd}$ and Shields number $\Theta$ to the following average properties of the transport layer: the average horizontal and vertical particle velocity $V_x=\overline{v_x}/\sqrt{(s-1)gd}$ and $V_z=\sqrt{\overline{v_z^2}}/\sqrt{(s-1)gd}$, respectively, and the characteristic transport layer thickness $Z=(\overline{z}-z_r)/d$ (roughly half of the actual thickness). Close to the threshold, two further equations allow to close the system and calculate $\Theta^r_t$: a relation between $V_x$ and $U_x$ and the law of the wall flow velocity profile, which is valid because transported particles do not disturb the flow in the limit $\Theta\rightarrow\Theta^r_t$. A final equation parametrizes the effect of cohesion on the cessation threshold by changing the boundary condition at the bed surface, where the intensity of cohesive forces is expressed in terms of the cohesion number $C=d^{-1}\sigma^{3/5}E^{-1/5}[(\rho_p-\rho_f)g]^{-2/5}$, with $\sigma$ the particle surface tension and $E$ the Young modulus (for quartz particles, $\sigma=3$~J/m$^2$ and $E=7\times10^{10}$~Pa). The full set of equations is
\begin{linenomath*}
\begin{subequations}
\begin{eqnarray}
 U_x-V_x&=&\left[\sqrt{\frac{1}{4}\sqrt[m]{\left(\frac{24}{C_d^\infty\mathrm{Ga}}\right)^2}+\sqrt[m]{\frac{4\mu_b}{3C_d^\infty}}}-\frac{1}{2}\sqrt[m]{\frac{24}{C_d^\infty\mathrm{Ga}}}\right]^m, \label{mub} \\
	V_z&=&\alpha\mu_b^{-1}\sqrt{V_x|_{\Theta=\Theta^r_t}V_x}\xrightarrow{\Theta\rightarrow\Theta^r_t}\alpha\mu_b^{-1}V_x, \label{afterclosure} \\
	Z&=&\beta\mu_b^{-1}\Theta+sV_z^2\xrightarrow{\Theta\rightarrow\Theta^r_t}\beta\mu_b^{-1}\Theta^r_t+sV_z^2, \label{Meanz} \\
	U_x&=&\sqrt{\Theta^r_t}f[\mathrm{Re}_d(Z+Z_\Delta),\mathrm{Re}_d]\equiv\sqrt{\Theta^r_t}f[\mathrm{Re}_{\overline{z}},\mathrm{Re}_d], \label{MeanU} \\
	V_x&=&\frac{2\sqrt{\Theta^r_t}}{\kappa}\sqrt{1-\exp\left[-\frac{1}{4}\gamma^2\kappa^2\left(U_x/\sqrt{\Theta^r_t}\right)^2\right]}, \label{VU} \\
	\mu_b&=&\mu^o_b\left[1+1.5(c_\mathrm{coh}C)^{5/3}\right], \label{mub2}
\end{eqnarray}
\end{subequations}
\end{linenomath*}
where $\mu^o_b=0.63$, $\alpha=0.18$, $\beta=0.9$, $\gamma=0.79$, and $Z_\Delta=0.7$ are model parameters obtained from adjusting equations~(\ref{mub}-\ref{VU}) to the cohesionless ($c_\mathrm{coh}=0$) transport simulations (Figures~\ref{AnalyticalModelValidation1} and \ref{AnalyticalModelValidation2}), $f$ a function associated with the law of the wall by \citet{GuoJulien07} [Appendix~\ref{LawWall} equation~(\ref{uxcomplex})], $\mathrm{Re}_d=\mathrm{Ga}\sqrt{\Theta^r_t}$ the particle Reynolds number, $\kappa=0.4$ is the von K\'arm\'an constant, and $c_\mathrm{coh}$ a dimensionless parameter quantifying the strength of adhesive forces ($c_\mathrm{coh}\ne0$ only for the dotted and dash-dotted lines in Figure~\ref{ModelvsMeasurements}).
In the supporting information, we provide a short and simple commented MATLAB code solving equations~(\ref{mub}--\ref{mub2}).

Equations~(\ref{mub}) and (\ref{afterclosure}) are obtained from the bed boundary condition: the ratio $\mu=-P_{zx}/P_{zz}$ between granular shear stress and pressure (the \textit{friction coefficient}) is relatively constant at the bed surface: $\mu_b\equiv\mu(z_r)=\mathrm{const}$, which links the vertical to the horizontal particle motion (Section~\ref{BedFriction}). In particular, the submerged weight of transported particles is linked to the horizontal fluid drag force, which is calculated using the empirical fluid drag law by \citet{Camenen07} that includes parameters accounting for the particle shape: $C^\infty_d=0.5$ and $m=2$ for spherical particles (used in Figures~\ref{AnalyticalModelValidation1} and \ref{AnalyticalModelValidation2} and for the dashed lines in Figure~\ref{ModelvsMeasurements}) and $C^\infty_d=1$ and $m=1.5$ for naturally-shaped particles (used for the nondashed lines in Figure~\ref{ModelvsMeasurements}). Equation~(\ref{Meanz}) takes into account that two mechanisms contribute to the characteristic transport layer thickness $Z$ (section~\ref{TransportLayerHeight}): intergranular contacts ($\beta\mu_b^{-1}\Theta$), which are significant in bedload transport, and the vertical motion of particles ($sV_z^2$). 

In contrast to equations~(\ref{mub}--\ref{Meanz}), equations~(\ref{MeanU}) and (\ref{VU}) require near-threshold conditions to be valid. Equation~(\ref{MeanU}) models the average fluid velocity $U_x$ near threshold conditions, like in previous studies \citep{Sauermannetal01,ClaudinAndreotti06,DuranHerrmann06,Kok10b,Pahtzetal12,Berzietal16,Berzietal17}, as the law of the wall velocity profile $\sqrt{\Theta^r_t}f[\mathrm{Re}_d(z-z_u)/d,\mathrm{Re}_d]$ evaluated at the mean transport layer height $z=Zd+z_r$, where we find from our direct transport simulations that the zero-velocity elevation $z_u$ is a slight constant distance below the effective location of particles when they rebound at the bed surface: $z_u=z_r-Z_\Delta d$. Equation~(\ref{VU}) is associated with a continuous rebound motion of individual particles along the bed surface near threshold conditions (section~\ref{HorizontalVelocity}).

Finally, equation~(\ref{mub2}), which is the only model equation that cannot be tested with the cohesionless transport simulations, describes the increase of bed particle resistance to shear stress, described in our model by the bed friction coefficient $\mu_b$, with the cohesion number $C$ due to adhesion forces, as derived in Appendix A.4 of \citet{ClaudinAndreotti06}. It results in an improved agreement of the analytical model with aeolian impact threshold measurements for conditions corresponding to small particles ($d<100\;\mu$m), for which the cohesion number $C$ significantly affects the model predictions when $c_\mathrm{coh}>0$ (Figure~\ref{ModelvsMeasurements}). However, note that the predictions for cohesionless conditions ($c_\mathrm{coh}=0$ and thus $\mu_b=\mu^o_b$) explain the majority of the measurements.

\subsubsection{Bed Boundary Condition} \label{BedFriction}
From our cohesionless transport simulations, we find that the cohesionless bed friction coefficient $\mu^o_b$ is relatively constant (Figure~S2). In the analytical model, we use $\mu^o_b=\mu_b(0)=0.63$, which is close to the simulation mean. The idea to describe steady sediment transport by a constant bed friction coefficient goes back to \citet{Bagnold56,Bagnold66,Bagnold73} and has been adopted in many analytical models of sediment transport \citep[e.g.,][]{AshidaMichiue72,EngelundFredsoe76,KovacsParker94,NinoGarcia94,NinoGarcia98a,Seminaraetal02,Parkeretal03,AbrahamsGao06,Sauermannetal01,DuranHerrmann06,Lammeletal12,Pahtzetal12,Pahtzetal13,Pahtzetal14,JenkinsValance14,Berzietal16,Berzietal17}. In a forthcoming paper \citep{PahtzDuran18b}, we show that the reason that this idea works so well is associated with continuous particle-bed rebounds rather than a granular yield criterion of dense granular flows \citep{Midi04}, as \citet{Bagnold56,Bagnold66,Bagnold73} originally proposed. Hence, $\mu_b=\mathrm{const}$ does not resemble a critical condition associated with the entrainment of bed sediment but rather the boundary condition that characterizes a continuous rebound state.

The bed friction coefficient $\mu_b$ is directly related to properties of the equilibrium transport layer. Indeed, using equations~(\ref{MomentumBalance}) and (\ref{HeightAverage}), we obtain equation~(\ref{mub}) via
\begin{linenomath*}
\begin{eqnarray}
 &\mu_b=-\frac{P_{zx}}{P_{zz}}(z_r)=-\frac{\int_{z_r}^\infty\rho\langle a_x\rangle\mathrm{d}z}{\int_{z_r}^\infty\rho\langle a_z\rangle\mathrm{d}z}=-\frac{\overline{a_x}}{\overline{a_z}}\simeq\frac{\overline{a^d_x}}{\tilde g}\simeq\frac{3\overline{C_d|\mathbf{u}-\mathbf{v}|(u_x-v_x)}}{4s\tilde gd}& \nonumber \\
 &\Rightarrow\mu_b=\frac{3}{4}C^{\mathrm{eff}}_d(U_x-V_x)^2\quad\text{with}\quad C^{\mathrm{eff}}_d=\left[\sqrt[m]{\frac{24}{\mathrm{Ga}(U_x-V_x)}}+\sqrt[m]{C_d^\infty}\right]^m& \nonumber \\
 &\Leftrightarrow U_x-V_x=\left[\sqrt{\frac{1}{4}\sqrt[m]{\left(\frac{24}{C_d^\infty\mathrm{Ga}}\right)^2}+\sqrt[m]{\frac{4\mu_b}{3C_d^\infty}}}-\frac{1}{2}\sqrt[m]{\frac{24}{C_d^\infty\mathrm{Ga}}}\right]^m,& 
\end{eqnarray}
\end{linenomath*}
where we approximated $-\overline{a_z}$ by the buoyancy-reduced gravity constant $\tilde g=(s-1)g/s$ because the mass balance [equation~(\ref{MassBalance})] implies that vertical fluid drag forces acting on ascending and descending particles tend to compensate each other \citep{Pahtzetal12}. Furthermore, we calculated $\overline{a_x}$ as the transport layer average of the horizontal fluid drag acceleration $a^d_x$ using the general drag law by \citet{Camenen07}, which gives rise to an effective average drag coefficient $C^{\mathrm{eff}}_d$. We also confirmed that the added mass contributions to both $\overline{a_z}$ and $\overline{a_x}$, which are potentially important when $s$ is close to unity, are negligible. The reason is the fact that the added mass force is proportional to the sum of the total noncontact force (as it would be in the absence of the added mass effect) and the total contact force, which tend to compensate each other on average when $s$ is close to unity.

We also use the bed boundary condition to derive an expression for the relationship between $V_z$ and $V_x$. First, we approximate $\mu=-P_{zx}/P_{zz}\simeq-P^t_{zx}/P^t_{zz}=-\langle v_zv_x\rangle/\langle v_z^2\rangle$, where the equality at the right-hand side follows from equations~(\ref{Pt}) and (\ref{MassBalance}). This approximation is trivial for saltation transport, for which $P_{ij}\simeq P^t_{ij}$. However, in a forthcoming paper \citep{PahtzDuran18b}, we show that it is approximately obeyed also for bedload transport even though $P^t_{ij}$ may be much smaller than $P_{ij}$. Hence, height averaging $\mu$ yields
\begin{linenomath*}
\begin{equation}
 \overline{\mu}\simeq-\frac{\overline{v_zv_x}}{\overline{v_z^2}}\simeq\frac{\overline{\langle v_x\rangle_\downarrow}-\overline{\langle v_x\rangle_\uparrow}}{\overline{\langle v_z\rangle_\uparrow}-\overline{\langle v_z\rangle_\downarrow}}\equiv2\alpha\frac{V_x}{(\overline{\langle v_z\rangle_\uparrow}-\overline{\langle v_z\rangle_\downarrow})/\sqrt{(s-1)gd}}, \label{beforeclosure}
\end{equation}
\end{linenomath*}
where $\langle\mathbf{v}\rangle_{\uparrow(\downarrow)}=\langle\mathbf{v}H[+(-)v_z]\rangle/\langle H[+(-)v_z]\rangle$, with $H$ the Heaviside function, is the average velocity of ascending (descending) particles and $\alpha$ a model parameter that is based on the assumption that $\overline{\langle v_x\rangle_{\uparrow}}\propto\overline{\langle v_x\rangle_{\downarrow}}$. However, this assumption only makes sense when the transport layer predominantly consists of particles rebounding at a nearly static bed surface, which limits its applicability to Shields numbers $\Theta$ sufficiently close to the threshold $\Theta^r_t$. In fact, when $\Theta$ is too large, the bed surface becomes fully mobile and a collisional layer of particles forms around it \citep{PahtzDuran17}, which contributes to the transport layer average. Finally, we approximate $\overline{\mu}(\overline{\langle v_z\rangle_\uparrow}-\overline{\langle v_z\rangle_\downarrow})\approx2\mu_b\sqrt{\overline{v_z^2}}$, yielding
\begin{linenomath*}
\begin{equation}
 V_z=\alpha\mu_b^{-1}V_x. \label{afterclosure2}
\end{equation}
\end{linenomath*}
Although this approximation significantly worsens the agreement with the simulation data (Figure~S3), it has the advantage of reducing the number of variables, which is necessary for closing the analytical model. Note that the simulation data indicate that $V_x$ in equations~(\ref{beforeclosure}) and (\ref{afterclosure2}) should be substituted by $\sqrt{V_x|_{\Theta=\Theta^r_t}V_x}$, resulting in equation~(\ref{afterclosure}), to be applicable to conditions far from the threshold $\Theta^r_t$ (Figures~\ref{AnalyticalModelValidation1}b and S2). In a future paper, we will show that this substitution follows from the energy balance. However, here we do not discuss it because the analytical threshold model only requires knowledge of the relationship between $V_z$ and $V_x$ in the limit $\Theta\rightarrow\Theta^r_t$, which is given by equation~(\ref{afterclosure2}).

\subsubsection{Characteristic Transport Layer Thickness} \label{TransportLayerHeight}
We obtain the characteristic transport layer thickness $Z$ from the vertical momentum balance of particles [equation~(\ref{MomentumBalance})] using equation~(\ref{HeightAverage}), partial integration, and $-a_z\simeq\tilde g$ via
\begin{linenomath*}
\begin{eqnarray}
 &Z=\frac{\overline{z}-z_r}{d}\simeq-\frac{\overline{(z-z_r)a_z}}{\tilde gd}=-\frac{\frac{1}{M_r}\int_{z_r}^\infty\rho(z-z_r)\langle a_z\rangle\mathrm{d}z}{\tilde gd}=-\frac{\frac{1}{M_r}\int_{z_r}^\infty(z-z_r)\frac{\mathrm{d}P_{zz}}{\mathrm{d}z}\mathrm{d}z}{\tilde gd}=\frac{\frac{1}{M_r}\int_{z_r}^\infty P_{zz}\mathrm{d}z}{\tilde gd}=\frac{\overline{P_{zz}/\rho}}{\tilde gd}& \nonumber \\
 &\Rightarrow Z=Z_c+sV_z^2,& \label{Meanz2}
\end{eqnarray}
\end{linenomath*}
where $Z_c=\overline{P^c_{zz}/\rho}/(\tilde gd)$ is the contribution to $Z$ from intergranular contacts and $sV_z^2=\overline{P^t_{zz}/\rho}/(\tilde gd)$ the contribution from the transport of particles between contacts [equation~(\ref{ParticleStress})].

The contribution $Z_c$ is associated with a collision-induced granular diffusion process. \citet{LeightonAcrivos86} showed that the thickness of such a diffusion layer scales with $\Theta d$, which was used to explain the asymptotic scaling $Q\propto\Theta^3$ of the viscous bedload transport rate (Figure~\ref{ModelComparisons1}a) \citep{CharruMouilleronArnould02,Charruetal16}. These findings indicate $Z_c\propto\Theta$.

Here we obtain nearly the same scaling using much simpler arguments. First, we assume that the predominant scale for the mean value of $P^c_{zz}/\rho$ is its value at the bed surface elevation $z_r$: $\overline{P^c_{zz}/\rho}\propto[P^c_{zz}/\rho](z_r)$ because the contact pressure $P^c_{zz}$ increases strongly with concentration $\rho$ \citep{ChialvoSundaresan13}, which maximizes near $z_r$. Second, we assume that intergranular contacts are only significant for bedload transport, in which collisional contributions to the particle stress tensor dominate kinetic contributions at $z_r$ \citep{PahtzDuran18b}: $[P^c_{zz}/\rho](z_r)\simeq[P_{zz}/\rho](z_r)=\mu_b^{-1}[-P_{zx}/\rho](z_r)$. Now, for intense bedload transport ($\Theta\gg\Theta^r_t$ and thus $-P_{zx}(z_r)/(\rho_p\tilde gd)\simeq\Theta$), a local granular rheology develops \citep{Maurinetal16}, in which a constant value $\mu_b=\mu(z_r)$ is associated with a constant value $\rho(z_r)/\rho_p$. For such conditions, we thus obtain
\begin{linenomath*}
\begin{equation}
 Z_c\propto\mu_b^{-1}\Theta. \label{Zc}
\end{equation}
\end{linenomath*}
Curiously, this scaling seems to hold fairly well even near the threshold $\Theta^r_t$, for which both $-P_{zx}(z_r)$ and $\rho(z_r)$ become small [$\rho(z_r)/\rho_p$ is no longer constant near $\Theta^r_t$ because the rheology becomes nonlocal for weak transport conditions \citep{PahtzDuran18b}].

\subsubsection{Horizontal Particle Velocity} \label{HorizontalVelocity}
Equation~(\ref{VU}), which seems to be generally valid near threshold conditions (Figure~\ref{AnalyticalModelValidation2}b), is foremost an empirical finding from our direct transport simulations as we have only been able to justify it under the drastic assumption that the particle motion near threshold conditions can be represented by the identical periodic hopping motion of individual particles along a perfectly flat bed (see below). Although this assumption is common in analytical sediment transport models \citep{JenkinsValance14,Berzietal16,Berzietal17}, it results in unphysical profiles of kinematic particle properties, such as a concentration profile $\rho(z)$ that is increasing with height $z$ \citep{PahtzDuran17}. It also rules out the possibility of intergranular contacts contributing to the transport layer properties, which are known to be crucial in bedload transport \citep{Schmeeckle14}. For example, the assumption implies that $Z_c=0$ and that the bed friction coefficient becomes a purely kinematic quantity associated with the lift-off ($\mathbf{v_\uparrow}$) and impact velocity $\mathbf{v_\downarrow}$ of the individual particles: $\mu_b=(v_{x\downarrow}-v_{x\uparrow})/(v_{z\uparrow}-v_{z\downarrow})$. For this reason, we only consider saltation transport and thus assume $Z=sV_z^2$ and $Z\gg Z_\Delta$ in the following analytical justification.

\noindent \textit{Saltation in log-layer velocity profile:} We consider the periodic hopping motion of a particle subjected to the log-layer velocity profile $u_x/\sqrt{(s-1)gd}=\kappa^{-1}\sqrt{\Theta}\ln[(z-z_u)/z_o]$, where $z_o=d/30$ is the surface roughness, implying $U_x\simeq\kappa^{-1}\sqrt{\Theta}\ln(30sV_z^2)$, which is the limit of equation~(\ref{MeanU}) for large $\mathrm{Re}_d$ using $Z\gg Z_{\Delta}$. Using equations~(\ref{mub}) and (\ref{afterclosure}), we can then write
\begin{linenomath*}
\begin{equation}
 \kappa^{-1}\sqrt{\Theta}\ln(30sV_z^2)-\alpha^{-1}\mu_bV_z=[U_x-V_x](\mu_b,\mathrm{Ga}).
\end{equation}
\end{linenomath*}
Now, neglecting the influence of fluid drag on the vertical particle motion implies that $V_z$ scales with the dimensionless vertical lift-off velocity $\tilde v_{\uparrow z}=v_{\uparrow z}/\sqrt{(s-1)gd}$:
\begin{linenomath*}
\begin{equation}
 \kappa^{-1}\sqrt{\Theta}\ln(30c_v^2s\tilde v_{\uparrow z}^2)-c_v\alpha^{-1}\mu_b\tilde v_{\uparrow z}=[U_x-V_x](\mu_b,\mathrm{Ga}), \label{TurbMin}
\end{equation}
\end{linenomath*}
where $c_v=V_z/\tilde v_{\uparrow z}$ is the scaling constant. Finally, we assume that $\Theta^r_t$ corresponds to the smallest value of $\Theta$ for which a solution of equation~(\ref{TurbMin}) for $\tilde v_{\uparrow z}$ exists: $\Theta^r_t=\min_{\tilde v_{\uparrow z}}\Theta(\tilde v_{\uparrow z})$, from which we obtain $\alpha^{-1}\mu_bV_z=c_v\alpha^{-1}\mu_b\tilde v_{\uparrow z}=2\kappa^{-1}\sqrt{\Theta^r_t}$ and thus
\begin{linenomath*}
\begin{equation}
 V_x=2\kappa^{-1}\sqrt{\Theta^r_t} \label{VxTurbLimit}
\end{equation}
\end{linenomath*}
using equation~(\ref{afterclosure}). Equation~(\ref{VxTurbLimit}) is, indeed, the limit of equation~(\ref{VU}) for large $U_x/\sqrt{\Theta^r_t}$ (i.e., small viscous sublayer of the turbulent boundary layer \citep{PahtzDuran17}).

\noindent \textit{Saltation in viscous velocity profile:} We consider the periodic hopping motion of a particle subjected to the viscous sublayer velocity profile $u_x/\sqrt{(s-1)gd}=\Theta\mathrm{Ga}(z-z_u)/d$, implying $U_x\simeq\Theta\mathrm{Ga}Z$, which is the limit of equation~(\ref{MeanU}) for small $\mathrm{Re}_d$ using $Z\gg Z_{\Delta}$. When sediment transport occurs within the viscous sublayer, Stokes drag usually dominates turbulent drag, meaning that equation~(\ref{mub}) can be approximated as its limit of small Galileo number $\mathrm{Ga}$: $U_x-V_x\simeq\mu_b\tilde v_s$, where $\tilde v_s=\mathrm{Ga}/18$ is the dimensionless terminal settling velocity. Hence, the same reasoning as in the log-layer case leads to [cf. equation~(\ref{TurbMin})]
\begin{linenomath*}
\begin{equation}
 18c_v^2\Theta s\tilde v_s\tilde v_{\uparrow z}^2-c_v\alpha^{-1}\mu_b\tilde v_{\uparrow z}=\mu_b\tilde v_s. \label{ViscMin}
\end{equation}
\end{linenomath*}
Equation~(\ref{ViscMin}) means that the larger $\tilde v_{\uparrow z}$ the smaller $\Theta$. Hence, the minimal value of $\Theta$ for which a solution of equation~(\ref{ViscMin}) exists is the one corresponding to the maximally possible value of $\tilde v_{\uparrow z}$. In fact, $\tilde v_{\uparrow z}$ cannot increase indefinitely as the magnitude of vertical velocity scales is limited by the settling velocity: $\tilde v^{\mathrm{max}}_{\uparrow z}\propto\tilde v_s$. Hence, the assumption $\Theta^r_t=\min_{\tilde v_{\uparrow z}}\Theta(\tilde v_{\uparrow z})$ leads to $\Theta^r_t\propto\mu_bs^{-1}\tilde v_s^{-2}$, which implies
\begin{linenomath*}
\begin{equation}
 V_x\propto U_x. \label{VxViscLimit}
\end{equation}
\end{linenomath*}
Equation~(\ref{VxViscLimit}) is, indeed, the limit of equation~(\ref{VU}) for small $U_x/\sqrt{\Theta^r_t}$ (i.e., large viscous sublayer \citep{PahtzDuran17}).

\subsection{Threshold Interpretation} \label{ThresholdInterpretation}
The analytical threshold model presented in section~\ref{AnalyticalModel} allows a physical interpretation of the threshold $\Theta^r_t$ defined by equation~(\ref{ThresholdDef}). In fact, two main ingredients of the analytical model are likely associated with a continuous rebound motion of particles: the constant bed friction coefficient $\mu_b$ (section~\ref{BedFriction}) and the relation between particle and fluid velocity (section~\ref{HorizontalVelocity}). In particular, the latter relation gives rise to interpret $\Theta^r_t$ as the minimal fluid shear stresses needed to compensate the average energy loss of rebounding particles by the fluid drag acceleration during particle trajectories, which is why we suggest to call $\Theta^r_t$ \textit{rebound threshold}. However, note that there are crucial differences between our analytical model and the previous rebound threshold models by \citet{Berzietal16,Berzietal17} (section~\ref{ContinuousReboundModels}).

The only regime in which this rebound threshold interpretation of $\Theta^r_t$ may be problematic is viscous bedload transport due to the absence of a hopping motion of particles (Section~\ref{BedloadSaltation}). However, because the analytical model also reproduces viscous bedload transport measurements (Figure~\ref{ModelvsMeasurements}), it may be necessary to widen the meaning of the term \textit{rebound} in this context to any kind of particle-bed interaction that dissipates energy.

\subsection{Physical Insights from the Analytical Threshold Model} \label{ModShields}
In this section, we shed light on the transition between bedload and saltation transport, present a general threshold diagram, and analyze the predictions of the analytical threshold model for specific conditions.

\subsubsection{Transition Between Bedload and Saltation Transport and General Threshold Diagram} \label{BedloadSaltation}
The only element of the analytical model that is associated with intergranular contacts and thus bedload transport is the quantity $Z_c=\beta\mu_b^{-1}\Theta$ in equation~(\ref{Meanz}), which describes the contribution of intergranular contacts to the characteristic transport layer thickness $Z$. Likewise, the parameter
\begin{linenomath*}
\begin{equation}
 S=\frac{Z-Z_c}{Z}=\frac{sV_z^2}{Z}<1 \label{S}
\end{equation}
\end{linenomath*}
describes the relative contribution of particle hops to $Z$. Consistent with the definitions given in section~\ref{Introduction}, one may thus quantify saltation transport (i.e., insignificant intergranular contacts) through $S\geq0.9$ (or any other value near but smaller than unity because $100\%$ saltation transport does not occur) and thus bedload transport (i.e., significant intergranular contacts) through $S<0.9$. 

\begin{figure}[htb]
 \begin{center}
  \includegraphics[width=1.0\columnwidth]{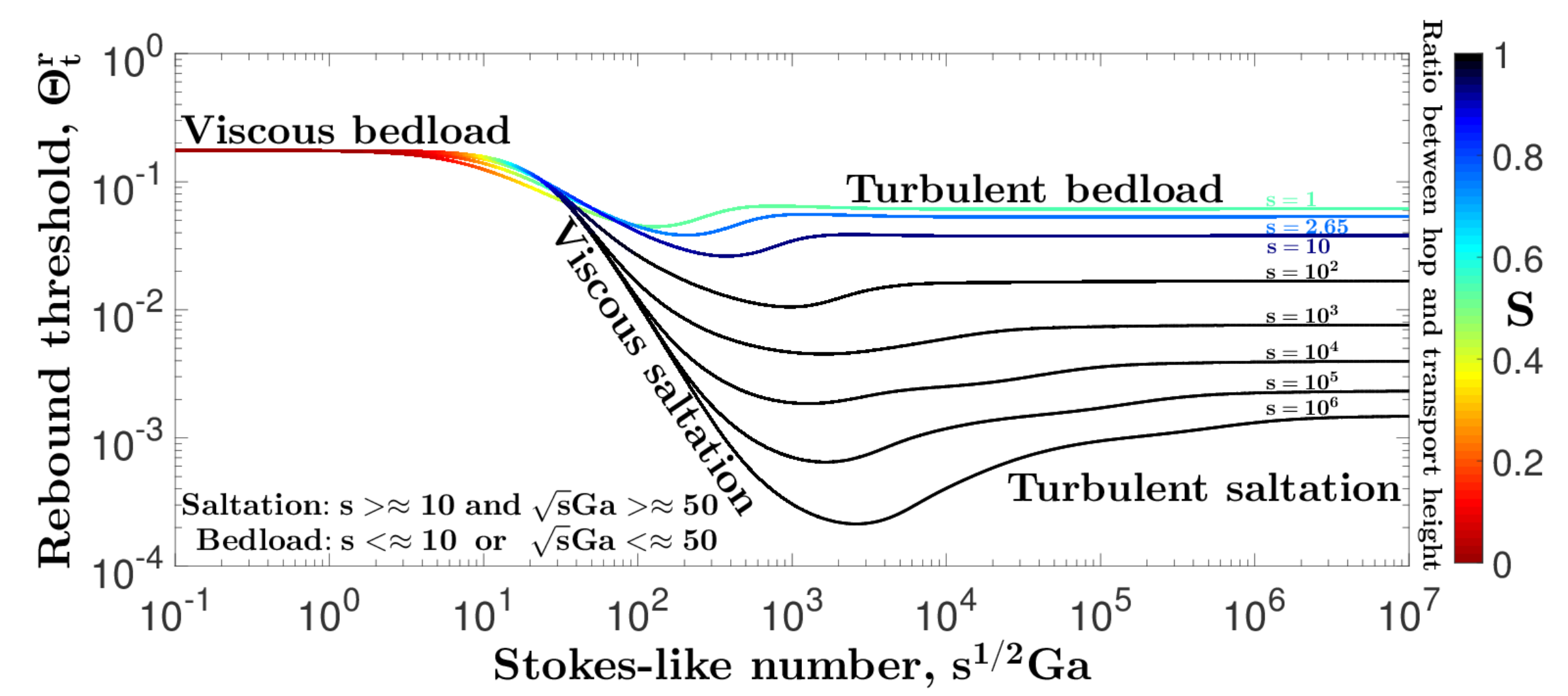}
 \end{center}
 \caption{{\bf General threshold diagram.} Rebound threshold $\Theta^r_t$ versus Stokes-like number $\sqrt{s}\mathrm{Ga}$ predicted by the analytical threshold model for cohesionless, naturally-shaped particles ($c_{\mathrm{coh}}=0$, $C^\infty_d=1$, $m=1.5$) and various particle-fluid-density ratios $s$. The color indicates the quantity $S$, which describes the relative contribution of particle hops to the characteristic transport layer height $Z$ [equation~(\ref{S})].}
 \label{ModifiedShieldsDiagram2}
\end{figure}

Figure~\ref{ModifiedShieldsDiagram2} shows the rebound threshold $\Theta^r_t$ versus the number $\sqrt{s}\mathrm{Ga}$ predicted by the analytical model for cohesionless particles and various particle-fluid-density ratios $s$ (solid lines), where the value of $S$ is indicated by the color of the line segments.
It can be seen that saltation transport occurs when the critical values $\sqrt{s}\mathrm{Ga}\approx50$ and $s\approx10$ are exceeded (which follows from the derivations in the next sections). Furthermore, it is apparent that particle hops are crucial in turbulent bedload transport (see also Movie~S2). In particular, for $s=2.65$ and large $\mathrm{Ga}$, the analytical model predicts $S=0.76$, which is consistent with measurements \citep[][their Figure~12]{Lajeunesseetal10}. The only transport regime that is truly dominated by intergranular contacts is viscous bedload transport ($\sqrt{s}\mathrm{Ga}\lesssim5$).

The dependent number $\sqrt{s}\mathrm{Ga}$ has previously been interpreted as a Stokes-like number \citep{Berzietal16,Berzietal17,Clarketal17}. Given that this number partially controls the transition between bedload and saltation transport, our interpretation of $\Theta^r_t$ as a rebound threshold, and the scaling $\Theta^r_t\propto(\sqrt{s}\mathrm{Ga})^{-2}$ for viscous saltation transport (see Figure~\ref{ModifiedShieldsDiagram2} and the derivation in the next section), it is much more meaningful to parametrize $\Theta^r_t$ by $\sqrt{s}\mathrm{Ga}$ than the particle Reynolds number $\mathrm{Re}_d$, as we did in the general threshold diagram (Figure~\ref{ModifiedShieldsDiagram2}).

\subsubsection{Threshold for Viscous Saltation Transport} \label{SaltationTransportRegimes}
We define viscous saltation transport as the regime in which the mean transport layer is large but nonetheless submerged within the viscous sublayer, and the fluid drag force is dominated by Stokes drag. We thus obtain $U_x-V_x\simeq\mu_b\mathrm{Ga}/18$ from equation~(\ref{mub}), $U_x\simeq\Theta^r_t\mathrm{Ga}Z$ from equations~(\ref{MeanU}) and (\ref{uxcomplex}), $V_x=\gamma U_x$ from equation~(\ref{VU}), and $Z\simeq\alpha^2\mu_b^{-2}sV_x^2$ from equations~(\ref{afterclosure}) and (\ref{Meanz}). Combined, these relations yield
\begin{linenomath*}
\begin{eqnarray}
 \Theta^r_t&\simeq&\frac{18\mu_b(1-\gamma)}{\alpha^2\gamma^2}(\sqrt{s}\mathrm{Ga})^{-2}, \label{Visc} \\
 Z&\simeq&\frac{\alpha^2\gamma^2}{324(1-\gamma)^2}(\sqrt{s}\mathrm{Ga})^2. \label{ZVisc}
\end{eqnarray}
\end{linenomath*}

\subsubsection{Threshold for Aerodynamically Smooth Turbulent Saltation Transport}
However, the transport height of particles cannot increase indefinitely within the viscous sublayer. At some point, the fastest moving particles exceed the viscous sublayer and become influenced by the buffer layer, though fluid drag remains viscous. Then $V_x$ becomes controlled by $\sqrt{\Theta^r_t}$ rather than $U_x$, which is encoded in the saturated value of equation~(\ref{VU}) (large $U_x/\sqrt{\Theta^r_t}$), reading $V_x\simeq2\kappa^{-1}\sqrt{\Theta^r_t}$. Furthermore, $U_x\lesssim\Theta^r_t\mathrm{Ga}Z$ because the viscous-sublayer profile represents an upper boundary for the mean fluid velocity profile in the buffer layer and log layer. Moreover, in the saturated regime, $U_x/V_x$ increases strongly (Figure~\ref{AnalyticalModelValidation2}b), and we thus approximate $U_x-V_x\simeq U_x$. These changes imply
\begin{linenomath*}
\begin{equation}
 \Theta^r_t\gtrsim\sqrt{\frac{\mu_b^3\kappa^2}{72\alpha^2}}\sqrt{s}^{-1}, \label{ViscTurb}
\end{equation}
\end{linenomath*}
which does not depend on $\mathrm{Ga}$.

\subsubsection{Threshold for Aerodynamically Rough Turbulent Saltation Transport}
For aerodynamically rough turbulent saltation transport, the viscous sublayer becomes negligible, and fluid drag tends to be in the turbulent regime. We thus obtain $U_x-V_x\simeq U_x=\sqrt{4\mu_b/(3C_d^\infty)}$ from equation~(\ref{mub}), $U_x\simeq\kappa^{-1}\sqrt{\Theta^r_t}\ln(Zd/z_o)=\kappa^{-1}\sqrt{\Theta^r_t}\ln(30Z)$ from equations~(\ref{MeanU}) and (\ref{uxcomplex}), $V_x=2\kappa^{-1}\sqrt{\Theta^r_t}$ from equation~(\ref{VU}), and $Z\simeq\alpha^2\mu_b^{-2}sV_x^2$ from equations~(\ref{afterclosure}) and (\ref{Meanz}). Combined, these relations yield
\begin{linenomath*}
\begin{eqnarray}
 \Theta^r_t&\simeq&\frac{4\kappa^2\mu_b}{3C_d^\infty\left[\ln\left(120\kappa^{-2}\alpha^2\mu_b^{-2}s\Theta^r_t\right)\right]^2}, \label{Turb} \\
 Z&\simeq&\frac{16\alpha^2s}{3C_d^\infty\mu_b\left[\ln\left(30Z\right)\right]^2}. \label{ZTurb}
\end{eqnarray}
\end{linenomath*}

\subsubsection{Minimum of the Rebound Threshold}
\begin{figure}[htb]
 \begin{center}
  \includegraphics[width=1.0\columnwidth]{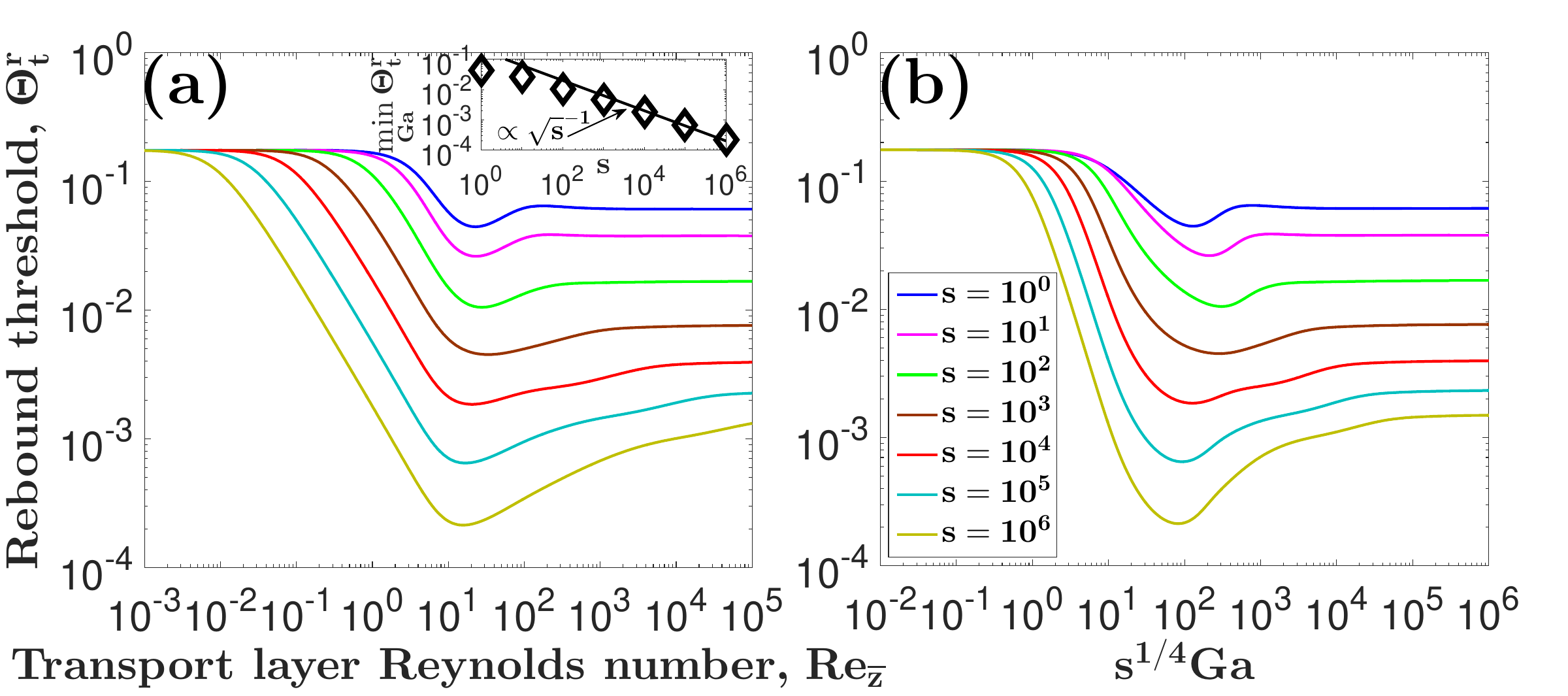}
 \end{center}
 \caption{{\bf Minimum of rebound threshold.} Rebound threshold $\Theta^r_t$ versus (a) transport layer Reynolds number $\mathrm{Re}_{\overline{z}}$ and (b) $s^{1/4}\mathrm{Ga}$ predicted by the analytical threshold model for cohesionless, naturally-shaped particles ($c_{\mathrm{coh}}=0$, $C^\infty_d=1$, $m=1.5$) and various particle-fluid-density ratios $s$. Inset: $\min\limits_{\mathrm{Ga}}\Theta^r_t(s,\mathrm{Ga})$ versus $s$.}
 \label{ModifiedShieldsDiagram1}
\end{figure}

Equation~(\ref{ViscTurb}) means that the decrease of $\Theta^r_t$ with $\mathrm{Ga}$ predicted by equation~(\ref{Visc}) must stop at some point, resulting in a local minimum $\min_{\mathrm{Ga}}\Theta^r_t(s,\mathrm{Ga})$. Hence, the right-hand side of equation~(\ref{ViscTurb}) is an estimate of this minimum value (inset of Figure~\ref{ModifiedShieldsDiagram1}a) and aerodynamically smooth turbulent saltation transport the regime that occurs around this minimum. Furthermore, because the minimum is associated with the mean transport layer height exceeding the viscous sublayer, it occurs at a critical value of the \textit{transport layer Reynolds number} $\mathrm{Re}_{\overline{z}}\equiv\mathrm{Re}_d(Z+Z_{\Delta})\equiv u^r_t(\overline{z}-z_u)/\nu$ (Figure~\ref{ModifiedShieldsDiagram1}a).
Using the relations valid for aerodynamically smooth turbulent saltation transport, this number can be approximated as
\begin{linenomath*}
\begin{equation}
 \mathrm{Re}_{\overline{z}}\simeq\frac{4\alpha^2s\Theta^r_t\sqrt{\Theta^r_t}\mathrm{Ga}}{\mu_b^2\kappa^2}\sim\mu_b^{1/4}s^{1/4}\mathrm{Ga}, \label{LocationMinimum}
\end{equation}
\end{linenomath*}
which is proportional to $s^{1/4}\mathrm{Ga}$ for cohesionless conditions ($\mu_b=\mu^o_b$). Indeed, Figure~\ref{ModifiedShieldsDiagram1}b shows that $s^{1/4}\mathrm{Ga}=200$ is a good predictor for the location of this minimum.

\section{Discussion} \label{Discussion}
In this section, we first support the main assumption of the analytical threshold model (section~\ref{SteadyState}). Then we discuss our results in the contexts of sediment entrainment (section~\ref{Thetate}) and transport intermittency (section~\ref{Intermittency}). Finally, we compare our analytical threshold model with previous ones (section~\ref{ThresholdComparison}).

\subsection{Steady State Assumption} \label{SteadyState}
The starting point of the analytical threshold model in Section~\ref{AnalyticalModel} is the assumption that, for any value of the Shields number $\Theta>\Theta^r_t$, there is a unique equilibrium (i.e., steady) transport state associated with continuous nonsuspended sediment transport that obeys equation~(\ref{ThresholdDef}). This assumption is not trivially fulfilled because it has been shown in previous numerical studies of bedload and saltation transport that transport becomes metastable when $\Theta$ falls below another critical value $\Theta^e_t$ that is larger than $\Theta^r_t$ \citep{Carneiroetal11,Clarketal15}: bulk transport will stop after an average time $t_s$, which obeys a power law $t_s-t^o_s\propto(\Theta^e_t-\Theta)^p$, where $t^o_s$ is a constant and $p<0$ the associated power. When bulk transport has stopped, it may be reinitiated by turbulent entrainment events \citep{Carneiroetal11}, leading to an intermittent transport regime in which $\Theta^e_t$ marks the transition to continuous transport, as discussed in section~\ref{Intermittency}.

\begin{figure}[htb]
 \begin{center}
  \includegraphics[width=1.0\columnwidth]{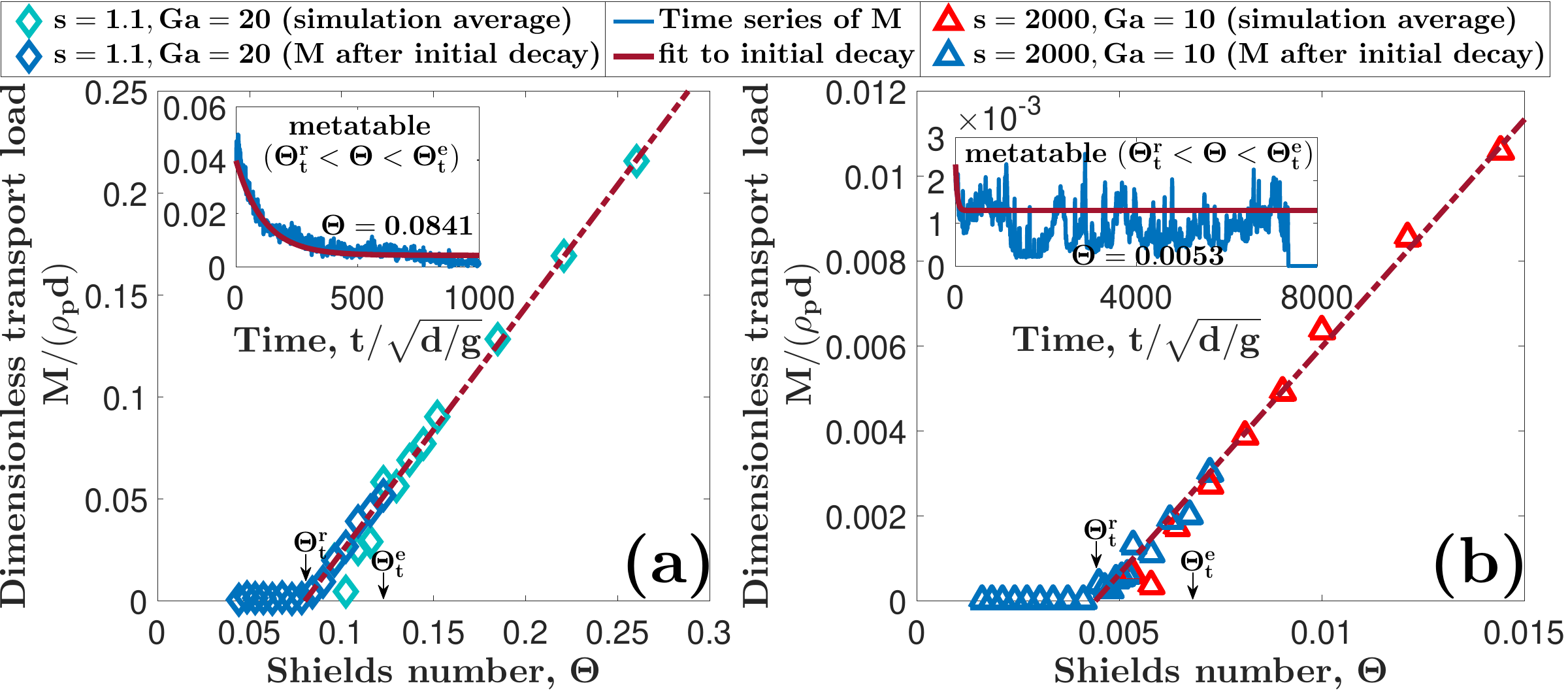}
 \end{center}
 \caption{{\bf Transported load: time series and the scaling at equilibrium.} Dimensionless transport load $M/(\rho_p d)$ versus Shields number $\Theta$ for (a) an example of turbulent bedload transport ($s=1.1$, $\mathrm{Ga}=20$) and (b) an example of saltation transport ($s=2,000$, $\mathrm{Ga}=10$). Insets: Decay time series of $M$ for metastable conditions ($\Theta^r_t<\Theta<\Theta^e_t$).}
\label{Extrapolation}
\end{figure}

In order to confirm that the metastable regime is nonetheless associated with an equilibrium state of continuous transport described by equation~(\ref{ThresholdDef}), we simulate the decay of the transport load $M=Q/\overline{v_x}^q$ for fixed combinations of $s$ and $\mathrm{Ga}$ and various Shields numbers $\Theta$ corresponding to both stable ($\Theta\geq\Theta^e_t$) and metastable ($\Theta<\Theta^e_t$) conditions, where the initial condition of this decay is given by the steady state of a stable simulation with larger $\Theta$ and where $\Theta^e_t$ is estimated as the smallest value of $\Theta$ at which short-time averages of $M$ always remain close to $M_e$ when running a simulation for a very long time. For these simulations, we test whether transport reaches, at least temporarily, a state that obeys equation~(\ref{ThresholdDef}): $M/(\rho_pd)\propto\Theta-\Theta^r_t$. The insets of Figure~\ref{Extrapolation} show the simulated decay behaviors exemplary for two typical conditions corresponding to turbulent bedload (Figure~\ref{Extrapolation}a) and saltation transport (Figure~\ref{Extrapolation}b).
It can be seen that, for any $\Theta>\Theta^r_t$, the system initially, indeed, decays toward an equilibrium transport state associated with the scaling $M_e/(\rho_pd)\propto\Theta-\Theta^r_t$ (dark blue symbols), which confirms that the steady state assumption of the analytical threshold model is physically meaningful. However, for $\Theta^r_t<\Theta<\Theta^e_t$, transport is on average (turquoise and red symbols) below the equilibrium state ($M<M_e$) because of the metastability of this state (insets of Figure~\ref{Extrapolation}). However, note that $M$ in Figure~\ref{Extrapolation}, although it can become very small, never completely vanishes, even when $\Theta\leq\Theta^r_t$, which is consistent with subsurface creeping \citep{Houssaisetal15}.

\subsection{Transport Metastability and Sediment Entrainment} \label{Thetate}
What is the mechanism responsible for the metastability of the equilibrium state of continuous transport for $\Theta^r_t<\Theta<\Theta^e_t$ described in section~\ref{SteadyState} as well as in previous studies \citep{Carneiroetal11,Clarketal15}. In the insets of Figure~\ref{Extrapolation}, it can be seen that an initial continuous decay is followed by discontinuous jumps afterward. According to the rebound threshold interpretation, the initial decay is due to the fact that the flow is oversaturated, carries too many particles, and therefore cannot compensate the average energy loss during an average particle rebound at the bed surface. However, as more and more particles are deposited, the particle-flow feedback becomes weaker and the flow stronger. The initial decay is over as soon as the flow is strong enough to compensate the energy losses during average rebounds. When this happens, deposition is no longer associated with the mean energy loss during a particle-bed rebound but with the deviation from this mean: a series of unfortunate rebounds caused by bed inhomogeneities that lead to deposition. In particular, once a rebounding particle does not reach a sufficiently large transport height, the flow is not able to compensate its mean energy loss during subsequent rebounds, and this particle will rapidly settle.

Such random losses of continuously rebounding particles must be compensated by the entrainment of bed sediment to ensure a stable steady state. In particular, particles must be entrained with a sufficiently high energy so that they can participate in the continuous rebound state. However, the rebound threshold interpretation of $\Theta^r_t$ does not involve quantitative statements about sediment entrainment because the analytical threshold model in section~\ref{AnalyticalModel} does not contain ingredients associated with sediment entrainment. According to this interpretation, there is thus no guarantee that sediment entrainment takes place when $\Theta$ exceeds $\Theta^r_t$. In fact, we hypothesize that insufficient sediment entrainment by particle-bed impacts for $\Theta^r_t<\Theta<\Theta^e_t$ is the reason why transport is metastable and that $\Theta^e_t$ is the impact entrainment threshold. To make our point, we focus on the cases $s=2,000$ and $\mathrm{Ga}=10$ (saltation transport in Earth's atmosphere) and $s=1.1$ and $\mathrm{Ga}=20$ (turbulent bedload transport), which are both fully sustained through entrainment by particle-bed impacts \citep{PahtzDuran17}. However, qualitatively similar statements to the ones we are about to make can also be made for all other fully impact-sustained conditions (i.e., $\mathrm{Im}=\mathrm{Ga}\sqrt{s+0.5}\gtrsim20$ \citep{PahtzDuran17}).

A number of studies experimentally investigated the physics of impact entrainment for a range of parameters that is typical for aeolian transport \citep{Mithaetal86,Werner90,Rioualetal00,Rioualetal03,Beladjineetal07,Ogeretal08,Ammietal09}. In such experiments, a particle is shot with varying velocity and angle onto a static particle bed and the outcome statistically analyzed. These studies report that the average energy $E_e$ of particles ejected by an impact is nearly independent of the velocity of the impactor, which instead mainly influences the number of ejected particles. This experimental observation has a crucial implication for the continuous rebound interpretation. If $E_e$ is too low compared with the energy $E_o$ required to participate in the continuous rebound state, ejected particles will lose more energy during their own subsequent rebounds than the fluid supplies during their short hops and thus rapidly settle. That is, ejected particles will not participate in the continuous rebound state, which means that there is no supply of continuously rebounding particles that can compensate random losses of them, and transport will thus eventually stop (i.e., metastability).

\begin{figure}[htb]
 \begin{center}
  \includegraphics[width=1.0\columnwidth]{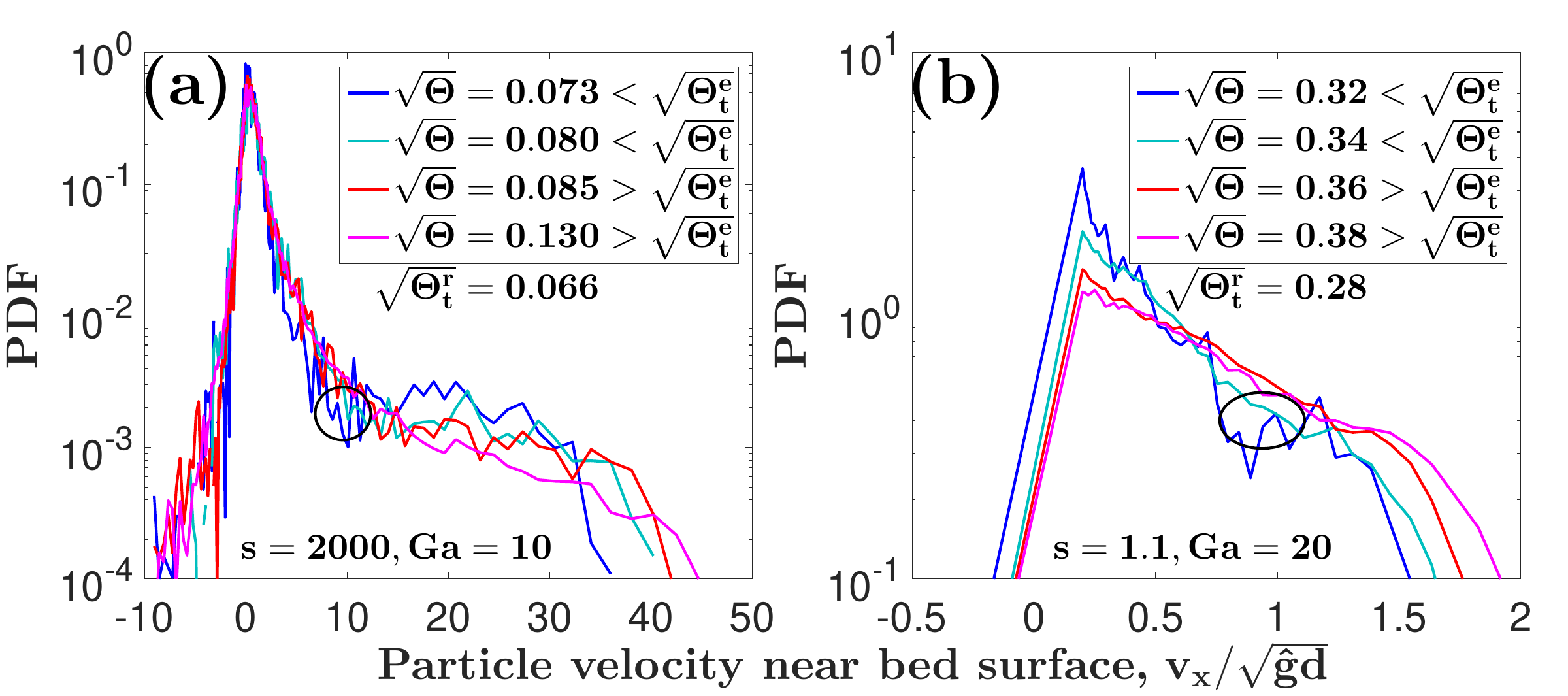}
 \end{center}
 \caption{{\bf Near-bed particle velocity distribution.} Distribution of the horizontal particle velocity $v_x/\sqrt{\hat gd}$ near the bed surface ($z\approx z_r$), where $\hat g=(s-1)g/(s+0.5)$ is the gravity constant reduced by the buoyancy and added-mass acceleration, for (a) an example of saltation transport ($s=2,000$, $\mathrm{Ga}=10$) and (b) an example of turbulent bedload transport ($s=1.1$, $\mathrm{Ga}=20$).}
 \label{Distribution}
\end{figure}

Indeed, this seems to be the case with our simulations when $\Theta^r_t<\Theta<\Theta^e_t$ (blueish lines) because of the bimodal character of the distribution of the horizontal particle velocity near the bed surface (Figure~\ref{Distribution}): a small peak at high velocity corresponding to continuous rebounders and a large peak at low velocity corresponding to entrained and creeping particles.
The gap for intermediate velocities (ellipses in Figure~\ref{Distribution}) means that bed particles only very rarely become continuous rebounders. In contrast, when $\Theta\geq\Theta^e_t$ (reddish lines in Figure~\ref{Distribution}), the peak at high velocity is flattened out, and the gap for intermediate velocities disappears. In the following, we propose two potential mechanisms, based on either isolated or cumulative particle-bed impacts, that may explain why the production rate of continuous rebounders strongly increases with $\Theta$ above $\Theta^r_t$.

\subsubsection{Mechanism Based on Isolated Particle-Bed Impacts} \label{Mechanism1}
For an isolated particle-bed impact to have a nonzero probability $P_r$ of directly promoting a bed particle to a continuous rebounder, the impacting particle must transmit more energy $\Delta E=E_\downarrow-E_\uparrow\equiv(1-e_b^2)E_\downarrow$, with $E_{\downarrow(\uparrow)}$ its energy before (after) impact and $e_b$ the bed restitution coefficient, to the bed than the minimum energy $E_o$ required to participate in the continuous rebound state: $\Delta E/E_o>1$. In particular, when $\Delta E$ is larger but close to $E_o$, small changes of $\Delta E$ have a relatively large effect on $P_r$.

The most violent impacts are those of the fastest continuous rebounders. For these particles, we expect $E_\uparrow=c_EE_o$, where $c_E>1$ is a proportionality constant of order unity. We thus obtain
\begin{linenomath*}
\begin{equation}
 \Delta E/E_o=c_E(e_b^{-2}-1).
\end{equation}
\end{linenomath*}
Measurements indicate $e_b\approx0.8$ for impact velocities and angles typical for aeolian saltation transport \citep{Beladjineetal07}, which suggests that $\Delta E/E_o$ is close to unity. Hence, the slight increase of $E_\downarrow$ with the Shields number $\Theta$ (cf. tails of the distributions in Figure~\ref{Distribution}), which is associated with a slight increase of $c_E$, may, indeed, be responsible for a relatively large increase of $P_r$ and thus explain why impact entrainment becomes sufficient at a value $\Theta^e_t$ that is larger than $\Theta^r_t$.

\subsubsection{Mechanism Based on Cumulative Particle-Bed Impacts} \label{Mechanism2}
When two transported particles in a sufficiently short sequence hit a bed particle, the second one has an increased probability of entraining and promoting it to a continuous rebounder as the bed particle does not fully recover from the first impact. This cumulative effect of successive particle-bed impacts, which has been neglected in most previous studies of impact entrainment \citep{AndersonHaff88,AndersonHaff91,HaffAnderson93,Rioualetal00,Rioualetal03,Ogeretal05,Ogeretal08,Beladjineetal07,Crassousetal07,Ammietal09,KokRenno09,ValanceCrassous09,Hoetal12,ComolaLehning17,Huangetal17,Tanabeetal17,Lammeletal17} and in numerous theoretical studies that link properties of aeolian saltation transport to the splash of isolated particle-bed impacts \citep{Andreotti04,Creysselsetal09,KokRenno09,Kok10b,Jenkinsetal10,Lammeletal12,Pahtzetal12,Huangetal14,JenkinsValance14,JenkinsValance18,WangZheng14,WangZheng15,Berzietal16,Berzietal17,Boetal17,LammelKroy17}, increases with the frequency of particle-bed impacts and thus with $\Theta-\Theta^r_t$ [equation~(\ref{ThresholdDef})]. Consistently, \citet{LeeJerolmack18}, who investigated bedload transport intermittency in a water flume as a function of the rate at which particles are fed at the flume entrance, reported a change in the velocity distribution of transported particles that is qualitatively very similar to the one shown in Figure~\ref{Distribution}: a bimodal distribution at low feeding rate (i.e., low impact frequency) turns into a unimodal distribution at high feeding rate (i.e., high impact frequency). In fact, when the characteristic time between impacts becomes smaller than the time the bed surface needs to recover from an impact, repeated impacts will increase the fluctuation motion of the bed. This impact-driven fluctuation motion weakens the links between neighboring bed surface particles when compared with a static bed. The weakest possible link of a bed surface particle to its neighbors is when it is about to detach from the bed. When the impactor hits such a particle, a fraction of the impact energy will be transferred to it but then not be further distributed among the many neighboring particles like for impacts on strongly linked static bed particles. Hence, the characteristic energy of entrained particles becomes coupled to the impact energy when a critical impact frequency is exceeded, which makes it much more likely that entrained particles participate in the continuous rebound state and thus may explain why impact entrainment becomes sufficient at a value $\Theta^e_t$ that is larger than $\Theta^r_t$.

\subsection{A New Conceptual Picture of Sediment Transport Intermittency} \label{Intermittency}
\begin{figure}[htb]
 \begin{center}
  \includegraphics[width=1.0\columnwidth]{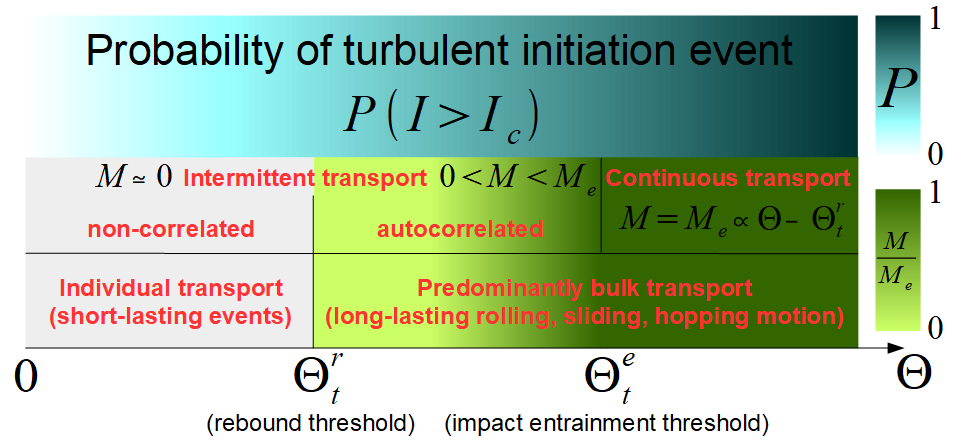}
 \end{center}
 \caption{{\bf Conceptual picture of sediment transport intermittency.} Three intermittency regimes are distinguished: (I) When $\Theta<\Theta^r_t$ (section~\ref{Regime1}), individual particles entrained by turbulent events ($I>I_c$) rapidly deposit again as the flow cannot compensate their average energy loss during collisions with the bed surface: the brief transport rate spike during such events interrupts an otherwise quiescent transport stage (transport load $M\simeq0$). (II) When $\Theta^r_t<\Theta<\Theta^e_t$ (section~\ref{Regime2}), the flow is sufficient in compensating such energy losses: transported particles tend to continuously rebound for comparably longer periods before depositing, resulting in a collective particle motion (i.e., bulk transport) and significant transport autocorrelations. Nonetheless, significant periods of rest still occur (i.e., transport cannot sustain equilibrium conditions: $M<M_e$) as turbulent entrainment events supply the transport layer with continuous rebounders only at an intermittent basis, while entrainment by particle bed-impacts is insufficient. (III) When $\Theta\geq\Theta^e_t$ (section~\ref{Regime3}), impact entrainment supplies the transport layer with continuous rebounders at a sufficiently high rate: transport is continuous and remains at equilibrium ($M=M_e$).}
 \label{SketchIntermittency}
\end{figure}

In this section, we put forward a new conceptual picture of sediment transport intermittency, which is sketched in Figure~\ref{SketchIntermittency}.
It combines the insights into the physics of sediment transport cessation from our study with the description of turbulent entrainment events by \citet{Diplasetal08}, \citet{Celiketal10}, and \citet{Valyrakisetal10,Valyrakisetal11,Valyrakisetal13}, which can supply the transport layer with new bed material on an intermittent basis. In a series of experimental and theoretical studies, these authors first showed that, for an entrainment event to occur, the instantaneous fluid force must exceed a critical value for a sufficiently long period of time so that the fluid impulse (i.e., the duration of an event with above-critical fluid force multiplied with the average value of this above-critical force) exceeds a threshold. \citet{Valyrakisetal13} then further refined this impulse criterion to an energy criterion: the time-integrated above-critical flow power $I$ \citep{Valyrakisetal13}, which is a measure for the mechanical energy transferred from flow to particle, must exceed a threshold for an entrainment event to occur: $I>I_c$. However, the probability of a successful energy event does depend not only on the Shields number $\Theta$ but also on the flow geometry and bed roughness, and the question of whether or not a particle-bed impact preceded the event \citep{Vowinckeletal16}, which is the reason why a critical Shields number criterion can only be seen as a crude approximation of this energy criterion and why visual measurements of the fluvial transport threshold scatter by more than an order of magnitude \citep{Valyrakisetal11}. In the subsequent sections, we will discuss the conceptual model and Figure 10 in more detail. 

\subsubsection{Intermittent Individual Transport ($\Theta<\Theta^r_t$)} \label{Regime1}
When the fluid shear stress $\Theta$ is below $\Theta^r_t$, individual particles entrained by turbulent events rapidly deposit again as the fluid shear stress cannot compensate their average energy loss when colliding with the bed surface according to our interpretation of $\Theta^r_t$ as a rebound threshold (section~\ref{ThresholdInterpretation}). This situation is typical for fluvial transport initiation studies that visually measure incipient motion using criteria based on critical amounts of individually moving particles, such as the criteria by \citet{Kramer35} (`weak' and `medium' motion). Consistently, these visually measured thresholds are typically smaller than the fluvial rebound threshold $\Theta^r_t$ measured using the reference method \citep{BuffingtonMontgomery97}.

\subsubsection{Intermittent Bulk Transport ($\Theta^r_t<\Theta<\Theta^e_t$)} \label{Regime2}
In contrast, when $\Theta^r_t<\Theta<\Theta^e_t$, particles entrained by turbulent events continuously rebound for comparably longer periods before they deposit again, giving rise to intermittent bulk transport and significant transport autocorrelations, as often observed in aeolian \citep{Lee87,StoutZobeck97,RasmussenSorensen99,Spiesetal00,BaasSherman05,Ellisetal12,PfeiferSchonfeldt12,Dupontetal13,Martinetal13,BauerDavidsonArnott14,Carneiroetal15,Shermanetal18} and fluvial systems \citep{HeathershawThorne85,Drakeetal88,Dinehart99,Anceyetal06,Anceyetal08,Heymanetal13,LeeJerolmack18}. The closer $\Theta$ comes to the impact entrainment threshold $\Theta^e_t$, the more such continuously rebounding particles the flow can carry, which increases the probability of impact entrainment events (section~\ref{Thetate}) and thus further enhances transport autocorrelation. Previous studies, indeed, showed that the transport autocorrelation function measured in fluvial bedload transport experiments cannot be explained by turbulence alone but that it must be a manifestation of an intrinsic transport mechanism that correlates sediment entrainment with the number of transported particles \citep{Anceyetal08,Anceyetal15,Heymanetal13}. However, as long as $\Theta<\Theta^e_t$, bulk transport will stop after a sufficient waiting period as impact entrainment is not sufficient to compensate random losses of continuous rebounders and since turbulent events only entrain particles on an intermittent basis. The resulting period of rest ends when a turbulent event reinitiates bulk transport.

\subsubsection{Continuous Bulk Transport ($\Theta\geq\Theta^e_t$)} \label{Regime3}
Once $\Theta$ exceeds $\Theta^e_t$, transport becomes continuous because impact entrainment events supply the transport layer with continuous rebounders at a sufficiently high rate, at which point the system remains at equilibrium and $M=M_e\propto\rho_pd(\Theta-\Theta^r_t)$. In contrast, as long as transport is intermittent (metastable), $M<M_e$. This conceptual picture implies that equation~(\ref{ThresholdDef}) is only valid for steady, continuous, but not for intermittent transport of nonsuspended sediment, which explains why bedload transport formulae like equation~(\ref{ThresholdDef}) fail when $\Theta\lesssim2\Theta^r_t$ \citep{MeyerPeterMuller48,Reckingetal12} and why one should therefore only use data representing bulk transport when testing such relations \citep{BunteAbt05,Singhetal09,ShihDiplas18}. In fact, measurements \citep{Carneiroetal15,MartinKok18a} and direct transport simulations \citep{Gonzalezetal17} suggest $\Theta^e_t\approx1.5\Theta^r_t$ for transport in Earth's atmosphere and $\Theta^e_t\approx2\Theta^r_t$ for transport in water, respectively.

If the picture described in the subsections above is accurate, it can be concluded that visual measurements that use the occurrence of intermittent bulk transport to characterize sediment transport cessation may result in threshold Shields numbers not far from $\Theta^r_t$. This provides a potential alternative means to experimentally estimate $\Theta^r_t$ for situations in which it is easy to distinguish individual from bulk transport. For example, bulk transport in aeolian environments is characterized by large particles hops, the occurrence of which can be relatively easily identified by visual inspection, which is what \citet{Bagnold37}and \citet{Chepil45} did when they measured the impact threshold. However, given that \citet{Bagnold37} and \citet{Chepil45} had no access to modern measurement equipment, such as video cameras, they probably overestimated the impact threshold and thus $\Theta^r_t$ (Figure~\ref{ModelvsMeasurements}). In contrast, the impact threshold that we estimated from the camera measurements by \citet{Carneiroetal15} seems to be more consistent with measurements of $\Theta^r_t$ (Figure~\ref{ModelvsMeasurements}).

\subsection{Comparison With Previous Threshold Models} \label{ThresholdComparison}
\subsubsection{Mean Flow Entrainment Models} \label{MeanFlowEntrainment}
Mean flow entrainment models derive a transport initiation threshold Shields number from a force balance, and/or torque balance, between mean fluid forces and resisting contact forces acting on a representative particle resting on the bed surface \citep{Bagnold36,Bagnold41,White40,Ward69,Iversenetal76,Schmidt80,Iversenetal87,IversenWhite82,WibergSmith87,Ling95,Dey99,Dey03,Lehningetal00,ShaoLu00,WuChou03,Luetal05,ClaudinAndreotti06,LucknerZanke07,VollmerKleinhans07,DeyPapanicolaou08,Recking09,Duranetal11,Leeetal12,Duanetal13,Bravoetal14,EdwardsNamikas15,AliDey16,Rousaretal16,Agudoetal17,Bravoetal17,HeOhara17}. Many of these models have been proposed to reproduce the Shields diagram, which displays two kinds of fluvial thresholds: a threshold obtained from extrapolating measurements of the transport rate to (nearly) vanishing transport and visual measurements of the initiation threshold of individual transport (see section~\ref{Intermittency} for details). The former threshold is the rebound threshold $\Theta^r_t$ according to our study, while the latter threshold is associated with much-larger-than-average instantaneous fluid forces during turbulent events \citep{Valyrakisetal11}. Hence, mean flow entrainment models do not capture the physics of the Shields diagram. However, mean flow entrainment models are able to capture the physics behind transport initiation in the absence of strong turbulent fluctuations, such as the initiation of laminar bedload transport, for which the predictions of certain mean flow entrainment models, indeed, match the experimental observations without any empirical input \citep{Agudoetal17}.

\subsubsection{Impact Entrainment Models}
In the aeolian transport community, a number of studies have proposed analytical models of equilibrium bulk saltation transport associated with impact entrainment, which take the dynamics of hopping particles into account \citep{Andreotti04,ClaudinAndreotti06,Kok10b,Lammeletal12,Pahtzetal12}. Such models approximate the turbulent flow field and the fraction $1-e_b^2$ of energy lost during particle-bed rebounds as their respective average values. Under these idealizations, the energy $E_\uparrow$ of a particle leaving the bed surface must be larger than a critical value $E_o$, which slightly depends on the lift-off angle, if it is to gain more energy $\Delta E=E_\downarrow-E_\uparrow$ during the following hop than it loses when successfully rebounding with the bed: $\Delta E>(1-e_b^2)E_\downarrow$ (cf. section~\ref{Mechanism1}). Otherwise, the particle will inevitably become trapped by the bed in subsequent hops even if the rebound probability $p_\mathrm{reb}$ is equal to unity. To our knowledge, all existing impact entrainment models effectively assume that, in equilibrium, all particles leave the surface with an above-critical lift-off energy, which is why the equilibrium condition in these models states that, on average, {\em{each hopping particle}} trapped by the bed due to rebound failure ($p_\mathrm{reb}<1$) must be compensated by a bed particle entrained by an impact of a hopping particle. In the limit of vanishing transport, this condition can be used to predict the cessation threshold $\Theta^e_t$ associated with impact entrainment \citep{ClaudinAndreotti06,Kok10b,Pahtzetal12}, which led to interesting insights into the physics of planetary saltation transport, such as the possibility of bulk saltation transport on Mars much below the bulk saltation transport initiation threshold \citep{Almeidaetal08,Kok10a}. However, the physics underlying these threshold models are strongly challenged by our study because the majority of bed particles acquire an energy that is significantly smaller than $E_o$ when entrained by particle-bed impacts (section~\ref{Thetate}), and these particles will inevitably become trapped by the bed regardless of the value of $p_\mathrm{reb}$. In fact, for the above idealizations of the turbulent flow field and particle-bed rebounds, the rebound threshold $\Theta^r_t$ is exactly the Shields number above which an equilibrium particle trajectory corresponding to continuously rebounding particles (i.e., particles that leave the surface with an above-critical energy) begins to exist according to our study. Hence, the equilibrium condition that impact entrainment models should use to obtain $\Theta^e_t$ should be as follows. On average, {\em{each continuously rebounding particle}} trapped by the bed due to rebound failure must be compensated by a bed particle that is entrained by a particle-bed impact {{\em and that acquires an above-critical lift-off energy upon entrainment}} (Section~\ref{Thetate}).

\subsubsection{Continuous Rebound Models} \label{ContinuousReboundModels}
Continuous rebound models of sediment transport cessation are a recent modeling technique put forward by \citet{Berzietal16,Berzietal17}. They represent the particle dynamics by particles moving in identical periodic trajectories over a perfectly flat bed, fix the impact and lift-off angles, and then look for the minimal Shields number that just allows a periodic identical trajectory. Although this picture conceptually agrees with our interpretation of $\Theta^r_t$ as a rebound threshold (section~\ref{ThresholdInterpretation}), there are crucial differences. Most importantly, by assuming identical periodic trajectories over a perfectly flat bed, the models by \citet{Berzietal16,Berzietal17} neglect the influences of intergranular contacts on the particle motion, which is the probable reason why these models overestimate measurements of the turbulent bedload transport threshold by an order of magnitude \citep[diamonds in][their Figure~7]{Berzietal16}. In contrast, intergranular contacts are accounted for in equation~(\ref{Meanz}) of our analytical threshold model. Furthermore, the models by \citet{Berzietal16,Berzietal17} assume that lubrication forces strongly damp particle-bed rebounds. In contrast, our numerical simulations indicate that lubrication is negligible in bedload transport (section~\ref{Lubrication}), which is why it is also neglected in the analytical model. Finally, in the idealized picture by \citet{Berzietal16,Berzietal17}, particles are never captured by the bed surface, which is why an upper bound on the impact velocity is imposed to prevent impact entrainment, the occurrence of which would not allow a steady state in the absence of particle capture. Hence, when the predicted threshold corresponds to a solution that exhibits this bounded value of the impact velocity, the continuous rebound threshold by \citet{Berzietal16,Berzietal17} essentially becomes an impact entrainment threshold. In contrast, according to our study, $\Theta^r_t$ is always a rebound threshold and the impact entrainment threshold $\Theta^e_t$ always larger than $\Theta^r_t$ because particle capture due to inhomogeneities of the bed surface is never sufficiently compensated by impact entrainment at $\Theta^r_t$ (section~\ref{Thetate}).

\section{Conclusions} \label{Conclusions}
In this study, we used numerical simulations that couple the discrete element method for the particle motion with a continuum Reynolds-averaged description of hydrodynamics to study the sediment transport threshold for a large range of Newtonian fluids driving transport. Among the various possible threshold definitions, we employed a definition that is applicable to arbitrary environmental conditions, insensitive to the presence of turbulent fluctuations around the mean turbulent flow, and meaningful in a geomorphological context: the indirect definition of the threshold $\Theta^r_t$ as the value of the Shields number $\Theta$ at which sediment transport rate relations like equation~(\ref{ThresholdDef}) predict vanishing transport (i.e., the reference method \citep{BuffingtonMontgomery97}). We then derived an analytical model to describe the simulation data. This model predicts $\Theta^r_t$ in arbitrary environments in agreement with available measurements in air and viscous and turbulent liquids despite not being fitted to these measurements (Figure~\ref{ModelvsMeasurements}). From the model and its interpretation, and the simulations, the following main conclusions can be drawn:
\begin{enumerate}
 \item The threshold $\Theta^r_t$ is a rebound threshold associated with the cessation of bulk sediment transport: the minimal fluid shear stress needed to compensate the average energy loss of rebounding particles by the fluid drag acceleration during particle trajectories. Hence, for $\Theta\geq\Theta^r_t$, transported particles continuously rebound for comparably longer periods before they deposit, whereas they deposit very quickly for $\Theta<\Theta^r_t$. For bedload transport, the normal restitution coefficient $e$ of binary particle collisions, whose effective value can be very small due to lubrication forces, does not significantly affect typical particle-bed rebounds and thus the value of $\Theta^r_t$ (section~\ref{Lubrication}).
 \item Neither $\Theta^r_t$ nor the average velocity $\overline{v_x}$ of transported particles (at least for Shields numbers $\Theta$ not too far from $\Theta^r_t$) depends significantly on the occurrence and statistics of bed sediment entrainment caused by the driving fluid and/or particle-bed impacts. That is, $\Theta^r_t$ is not related to the balance of forces and/or torques, including those induced by particle-bed impacts, acting on particles resting on the bed surface.
 \item The reference method and the visual method used to determine fluvial transport thresholds displayed in the Shields diagram \citep{BuffingtonMontgomery97} are fundamentally different: the former method results in the rebound threshold $\Theta^r_t$ for bulk motion, whereas the latter method results in a smaller threshold associated with incipient motion of individual particles due to turbulent events \citep{Valyrakisetal11}.
 \item For a given particle-fluid-density ratio $s$, the minimum value of $\Theta^r_t$ as a function of the Galileo number $\mathrm{Ga}$, or the particle Reynolds number $\mathrm{Re}_d$, occurs when transported particles begin to exceed the viscous sublayer of the turbulent boundary layer (i.e.,	when $s^{1/4}\mathrm{Ga}\approx200$, see Figure~\ref{ModifiedShieldsDiagram1}).
 \item The transition between bedload and saltation transport is characterized by critical values of two dimensionless numbers. Saltation transport occurs when $\sqrt{s}\mathrm{Ga}\gtrsim50$ and $s\gtrsim10$, whereas bedload transport occurs when $\sqrt{s}\mathrm{Ga}\lesssim50$ or $s\lesssim10$, which suggests that one should use the Stokes-like number $\sqrt{s}\mathrm{Ga}$ as the dependent parameter of a general Shields-like threshold diagram (Figure~\ref{ModifiedShieldsDiagram2}).
 \item The threshold Shields number $\Theta^e_t$ at which entrainment by particle-bed impacts can fully compensate the rare, yet occurring deposition of continuously rebounding particles is larger than $\Theta^r_t$ (for aeolian saltation transport, $\Theta^e_t\approx1.5\Theta^r_t$, for subaqueous bedload transport, $\Theta^e_t\approx2\Theta^r_t$ \citep{Carneiroetal15,Gonzalezetal17,MartinKok18a}). We have proposed two potential mechanisms to explain this inequality (section~\ref{Thetate}). One mechanism is the cumulative effect of successive particle-bed impacts, which causes a fluctuation motion of bed surface particles that makes them more susceptible for entrainment in a continuous rebound state. Note that recent numerical and experimental studies indicate that impact entrainment dominates direct fluid entrainment in turbulent bedload transport \citep{Vowinckeletal16,PahtzDuran17,LeeJerolmack18}.
 \item The impact entrainment threshold $\Theta^e_t$ also marks the transition between intermittent and continuous transport. Below $\Theta^e_t$, turbulent events sustain transport on an intermittent basis, where the intermittency characteristics depend on whether the Shields number $\Theta$ is above or below the rebound threshold $\Theta^r_t$ (Figure~\ref{SketchIntermittency}).
 \item Sediment transport rate relations like equation~(\ref{ThresholdDef}) are only applicable to steady, continuous transport of nonsuspended sediment (i.e., $\Theta\geq\Theta^e_t$) but fail to describe intermittent transport (i.e., $\Theta<\Theta^e_t$), which explains why bedload transport formulae like equation~(\ref{ThresholdDef}) fail when $\Theta\lesssim2\Theta^r_t$ \citep{MeyerPeterMuller48,Reckingetal12} and why one should therefore only use data representing bulk transport when testing such relations \citep{BunteAbt05,Singhetal09,ShihDiplas18}.
\end{enumerate}
Finally, we would like to note that we have exclusively considered a nearly horizontal sediment bed, consisting of particles of relatively uniform size [$d_p\in(0.8d,1.2d)$, evenly distributed], submerged in a deep turbulent boundary layer in this study. Unraveling the dependencies of the different transport thresholds on the grain size polydispersity, boundary layer height (e.g., for supercritical water flows), and bed slope remains a major challenge \citep{Wilcock93,Wilcock98,WilcockCrowe03,Lambetal08,Prancevicetal14,PrancevicLamb15,Maurinetal18,Seiletal18}.

\appendix
\section{Law of the Wall [Function $f$ in equation~(\ref{MeanU})]} \label{LawWall}
The law of the wall by \citet{GuoJulien07} is
\begin{linenomath*}
\begin{eqnarray}
 \frac{u_x}{\sqrt{(s-1)gd}} & = & \sqrt{\Theta^r_t}f\left( \mathrm{Re}_d(z-z_u)/d,\mathrm{Re}_d \right), \nonumber \\
 f\left( \mathrm{Re}_dz/d,\mathrm{Re}_d \right) & = & 7\arctan\left( \frac{\mathrm{Re}_d}{7}\frac{z}{d} \right)+\frac{7}{3}\arctan^3 \left( \frac{\mathrm{Re}_d}{7}\frac{z}{d} \right) \nonumber \\ 
 && - 0.52\arctan^4\left( \frac{\mathrm{Re}_d}{7}\frac{z}{d} \right) + \ln\left[1+\left(\frac{\mathrm{Re}_d}{B}\frac{z}{d}\right)^{(1/\kappa)}\right] \nonumber \\
 &&-\frac{1}{\kappa}\ln\left\{1+0.3\mathrm{Re}_d \left[1-\exp\left(-\frac{\mathrm{Re}_d}{26}\right)\right]\right\}, \label{uxcomplex}
\end{eqnarray}
\end{linenomath*}
where $\kappa=0.4$ and $B=\exp(16.873\kappa-\ln9)$. This version of the law of the wall has the advantage of providing a single equation for all flow regimes. Within the viscous sublayer of the turbulent boundary layer $u_x/\sqrt{(s-1)gd} \rightarrow \Theta^r_t\mathrm{Ga}(z-z_u)/d$, whereas in the log-layer $u_x/\sqrt{(s-1)gd} \rightarrow \kappa^{-1}\sqrt{\Theta^r_t}\ln[(z-z_u)/z_o]$. The roughness length $z_o$ equals $d/(9\mathrm{Re}_d)$ in the hydraulically smooth and $d/30$ in the hydraulically rough regime \citep{GuoJulien07}.

\acknowledgments
All data used in this study are reported in the figures and cited literature. A short MATLAB code calculating the threshold is available in the supporting information. We thank Dr. Rapha\"el Maurin for providing us data from his numerical sediment transport model. We thank Dr. Manousos Valyrakis and two anonymous reviewers for critically and constructively reviewing this paper. We acknowledge support from grant National Natural Science Foundation of China (No.~11750410687, No. 11550110179, and No.~91647209).

\begin{notation}
 \notation{$\tau$} Fluid shear stress [Pa]
 \notation{$\rho_p$} Particle density [kg/m$^3$]
 \notation{$\rho$} Particle mass density [kg/m$^3$] (particle density $\times$ volume fraction)
 \notation{$\rho_f$} Fluid density [kg/m$^3$]
 \notation{$g$} Gravitational constant [m/s$^2$]
 \notation{$\tilde g=(1-\rho_f/\rho_p)g$} Buoyancy-reduced gravitational constant [m/s$^2$]
 \notation{$d$} Characteristic particle diameter [m]
 \notation{$\nu$} Kinematic viscosity [m$^2$/s]
 \notation{$E$} Young modulus [Pa] (for quartz particles, $E=7\times10^{10}$~Pa)
 \notation{$\sigma$} Particle surface tension [J/m$^2$] (for quartz particles, $\sigma=3$ J/m$^2$)
 \notation{$x$} Horizontal coordinate in the flow direction parallel to the bed [m]
 \notation{$z$} Vertical coordinate in the direction normal to the bed [m]
 \notation{$z_r$} Bed surface elevation [m] (effective location of energetic particle-bed rebounds)
 \notation{$z_u$} Zero-fluid-velocity elevation [m]
 \notation{$z_o$} Roughness length [m]
 \notation{$\overline{z}$} Transport layer average of $z$ [m]
 \notation{$\overline{v_x}$} Transport layer average of the horizontal particle velocity [m/s]
 \notation{$\overline{v_x}^q$} Flux-weighted average of the horizontal particle velocity [m/s]
 \notation{$\overline{a_x}$} Transport layer average of the horizontal particle acceleration [m/s]
 \notation{$\overline{a_z}$} Transport layer average of the vertical particle acceleration [m/s]
 \notation{$Q$} Sediment transport rate [kg/(ms)]
 \notation{$Q_r$} Sediment transport rate above the bed surface [kg/(ms)]
 \notation{$M_r=Q_r/\overline{v_x}$} Layer-based transport load [kg/m$^2$]
 \notation{$M=Q/\overline{v_x}^q$} Flux-based transport load [kg/m$^2$]
 \notation{$P_{ij}=P^t_{ij}+P^c_{ij}$} Particle stress tensor [Pa] (shear stress $-P_{zx}$, pressure $P_{zz}$)
 \notation{$P^t_{ij}$} Kinetic contribution to particle stress tensor [Pa]
 \notation{$P^c_{ij}$} Contact contribution to particle stress tensor [Pa]
 \notation{$\Theta=\tau/((\rho_p-\rho_f)gd)$} Shields number
 \notation{$\Theta^r_t$} Shields number at the cessation threshold of bulk transport (`rebound threshold')
 \notation{$\Theta^e_t$} Shields number at the impact entrainment threshold (= continuous transport threshold)
 \notation{$s=\rho_p/\rho_f$} Particle-fluid-density ratio
 \notation{$\mathrm{Ga}=\sqrt{(s-1)gd^3}/\nu$} Galileo number
 \notation{$\mathrm{Re}_d=\mathrm{Ga}\sqrt{\Theta^r_t}$} Particle Reynolds number
 \notation{$\mathrm{Re}_{\overline{z}}=\mathrm{Re}_d(Z+Z_\Delta)$} Transport layer Reynolds number
 \notation{$C=d^{-1}\sigma^{3/5}E^{-1/5}((\rho_p-\rho_f)g)^{-2/5}$} Cohesion number
 \notation{$\kappa=0.4$} von K\'arm\'an constant
 \notation{$C_d$} Drag coefficient
 \notation{$C^\infty_d$} Turbulent drag coefficient (for natural particles, $C^\infty_d=1$)
 \notation{$m$} Parameter associated with drag law by \citet{Camenen07} (for natural particles, $m=1.5$)
 \notation{$Z=(\overline{z}-z_r)/d$} Dimensionless characteristic transport layer thickness
 \notation{$Z_\Delta=(z_r-z_u)/d=0.7$} Dimensionless distance between $z_u$ and $z_r$ (model parameter)
 \notation{$Z_c=\overline{P^c_{zz}/\rho}/(\tilde gd)$} Contribution of particle contacts to $Z$ (important for bedload transport)
 \notation{$U_x=\overline{u_x}/\sqrt{(s-1)gd}$} Dimensionless average horizontal fluid velocity
 \notation{$V_x=\overline{v_x}/\sqrt{(s-1)gd}$} Dimensionless average horizontal particle velocity
 \notation{$V_z=\overline{\sqrt{v_z^2}}/\sqrt{(s-1)gd}$} Dimensionless average vertical particle velocity
 \notation{$\mu=-P_{zx}/P_{zz}$} Ratio between particle shear stress and pressure (`friction coefficient')
 \notation{$\mu_b=\mu(z_r)$} Bed friction coefficient
 \notation{$\mu^o_b=0.63$} Cohesionless bed friction coefficient (model parameter)
 \notation{$c_\mathrm{coh}$} Model parameter quantifying the strength of adhesive forces
 \notation{$\alpha=0.18$} Model parameter linking horizontal speed of ascending and descending particles
 \notation{$\beta=0.9$} Model parameter associated with $Z_c$ (important for bedload transport)
 \notation{$\gamma=0.79$} Model parameter associated with the link between $V_x$ and $U_x$
 \notation{$f$} Function associated with the law of the wall by \citet{GuoJulien07} [equation~(\ref{uxcomplex})]
 \notation{$S=(Z-Z_c)/Z$} Relative contribution of particle hops to $Z$
 \notation{$I$} Time-integrated flow power (only events with above-critical flow power are considered)
 \notation{$I_c$} Critical value of $I$ above which a turbulent event leads to bed particle entrainment
\end{notation}

\section*{Movie S1.} 

Time evolution of the simulated particle-fluid system for $s=2000$, $\mathrm{Ga}=20$, $\Theta\simeq2\Theta^r_t$, and weakly damped interparticle collisions (restitution coefficient $e=0.9$). The flow velocity is shown as a background color with warm colors corresponding to high velocities and cold colors to small velocities. The horizontal and vertical axes are measured in mean particle diameters. Only $1/4$ of the simulated horizontal domain is shown, which is why particles occasionally enter the system from the left. It can be seen that entrainment by particle-bed impacts dominates entrainment by the mean turbulent flow (fluctuations around the mean are neglected), and that the flow maintains a continuous rebound motion of transported particles.

\section*{Movie S2.} 

Time evolution of the simulated particle-fluid system for $s=2.65$, $\mathrm{Ga}=20$, $\Theta\simeq1.5\Theta^r_t$, and nearly maximally damped interparticle collisions (restitution coefficient $e=0.01$). The flow velocity is shown as a background color with warm colors corresponding to high velocities and cold colors to small velocities. The horizontal and vertical axes are measured in mean particle diameters. Only $1/4$ of the simulated horizontal domain is shown, which is why particles occasionally enter the system from the left. It can be seen that entrainment by particle-bed impacts dominates entrainment by the mean turbulent flow (fluctuations around the mean are neglected), and that the flow maintains a continuous rebound motion of transported particles despite $e=0.01$.

\section*{Movie S3.} 

Time evolution of the simulated particle-fluid system for $s=2000$, $\mathrm{Ga}=20$, $\Theta\simeq3\Theta^r_t$, and nearly maximally damped interparticle collisions (restitution coefficient $e=0.01$). The flow velocity is shown as a background color with warm colors corresponding to high velocities and cold colors to small velocities. The horizontal and vertical axes are measured in mean particle diameters. Only $1/4$ of the simulated horizontal domain is shown, which is why particles occasionally enter the system from the left. It can be seen that entrainment by particle-bed impacts dominates entrainment by the mean turbulent flow (fluctuations around the mean are neglected), and that particles move in large hops despite $e=0.01$.

\section*{MATLAB code.}

Commented, self-explaining MATLAB code to compute the transport threshold $\Theta^r_t$ from the analytical model equations.

\setcounter{figure}{0}
\renewcommand{\thefigure}{S\arabic{figure}}

\begin{figure}[htbp]
 \begin{center}
  \includegraphics[width=1.0\columnwidth]{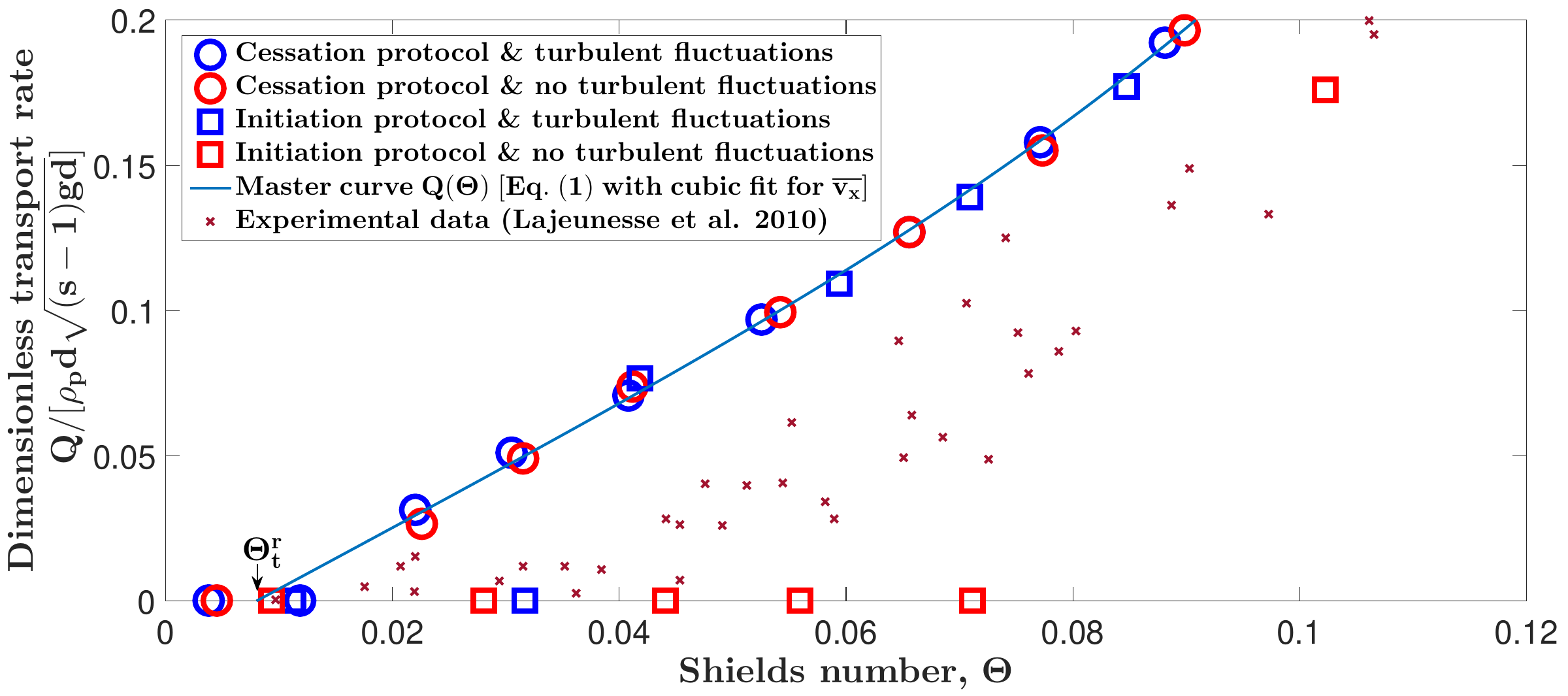}
 \end{center}
 \caption{Dimensionless sediment transport rate $Q/[\rho_pd\sqrt{(s-1)gd}]$ versus Shields number $\Theta$. Blue (red) symbols correspond to numerical sediment transport simulations with (without) turbulent fluctuations around the mean turbulent flow. Circles correspond to numerical sediment transport simulations using an initially mobile bed (steady transport), squares to an initially static bed, and crosses to measurements by \citet{Lajeunesseetal10}. This figure is an extension of Figure~6 of \citet{Maurinetal15} with mobile bed data. Measurements and simulations, though quantitatively different, exhibit a similar qualitative behavior, and it is the sole purpose of this figure to estimate the qualitative effect of turbulent fluctuations and the numerical protocol on the transport rate data.}
%
\end{figure}
\begin{figure}[htbp]
 \begin{center}
  \includegraphics[width=1.0\columnwidth]{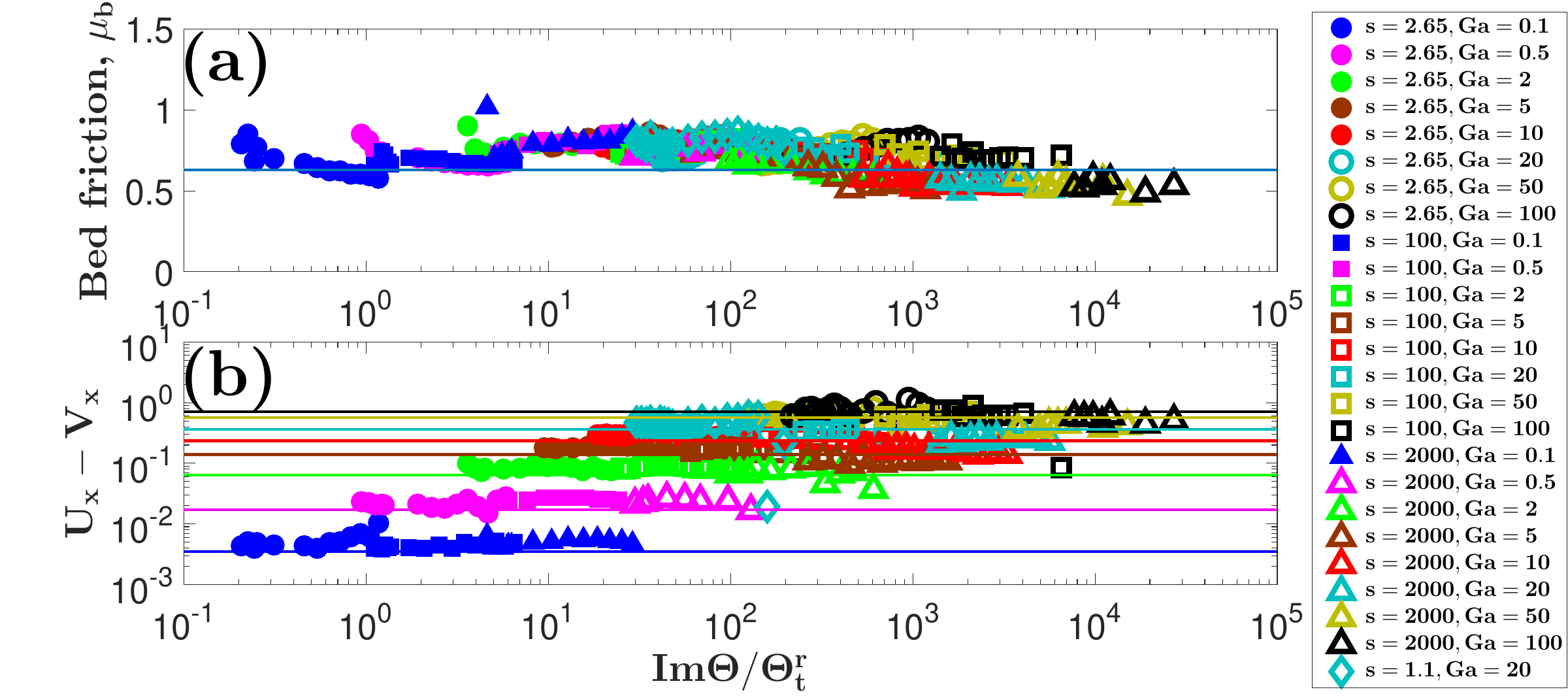}
 \end{center}
 \caption{(a) Bed friction coefficient $\mu_b$ and (b) dimensionless fluid-particle-velocity difference $U_x-V_x$ versus product of impact number $\mathrm{Im}=\mathrm{Ga}\sqrt{s+0.5}$ and $\Theta/\Theta^r_t$. Symbols correspond to data from our cohesionless transport simulations ($c_{\mathrm{coh}}=0$) for varying $s$, $\mathrm{Ga}$, and $\Theta$. The solid line in (a) corresponds to the constant value $\mu^o_b=0.63$ used in the analytical model. The colored solid lines in (b), each color corresponding to the same value of $\mathrm{Ga}$ as the likewise colored symbols, show $U_x-V_x$ calculated by the analytical model.}
\end{figure}
\begin{figure}[htbp]
 \begin{center}
  \includegraphics[width=1.0\columnwidth]{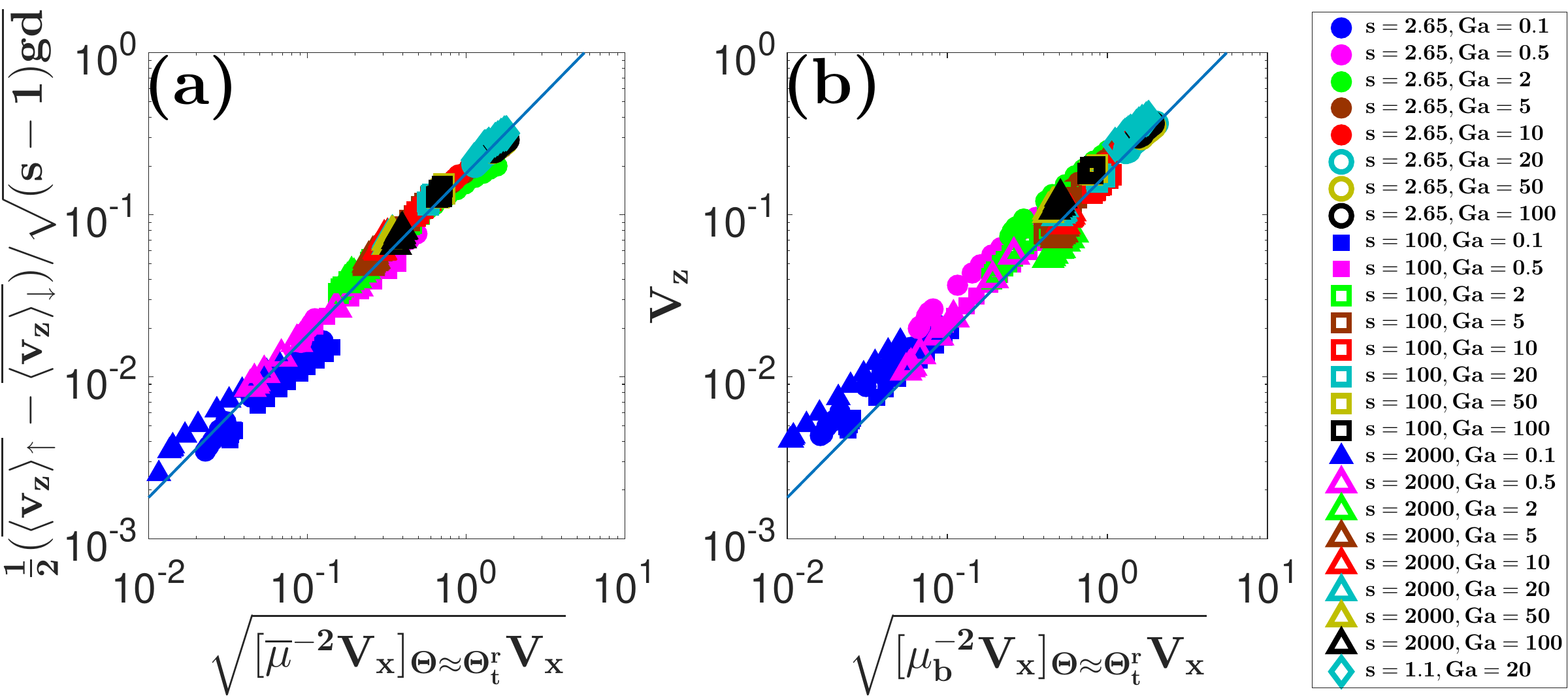}
 \end{center}
 \caption{(a) $\frac{1}{2}(\overline{\langle v_z\rangle_{\uparrow}}-\overline{\langle v_z\rangle_{\downarrow}})/\sqrt{(s-1)gd}$ versus $\sqrt{[\overline{\mu}^{-2}V_x]_{\Theta\approx\Theta^r_t}V_x}$ and (b) $V_z \equiv \sqrt{\overline{v^2_z}}/\sqrt{(s-1)gd}$ versus $\sqrt{[\mu_b^{-2}V_x]_{\Theta\approx\Theta^r_t}V_x}$. Symbols correspond to data from our cohesionless transport simulations ($c_{\mathrm{coh}}=0$) for varying $s$, $\mathrm{Ga}$, and $\Theta$. Simulated values of $\mu_b$ and $\overline{\mu}$ are used. The solid lines show the proportionality between abscissa and ordinate using a proportionality factor of $\alpha=0.18$.}
\end{figure}

\newpage


\end{document}